# A unified model for the human response to lipopolysaccharide-induced inflammation

Kristen A. Windoloski, Elisabeth O. Bangsgaard, Atanaska Dobreva, Johnny T. Ottesen, and Mette S. Olufsen

**Abstract** This study develops a unified model predicting the whole-body response to endotoxin. We simulate dynamics using differential equations examining the response to a lipopolysaccharide (LPS) injection. The model tracks pro- and anti-inflammatory cytokines (TNF-$\alpha$, IL-6, IL-10), concentrations of corticotropin-releasing hormone (CRH), adrenocorticotropic hormone (ACTH), and cortisol in the hypothalamic-pituitary-adrenal (HPA) axis. Daily hormonal variations are integrated into the model by including circadian oscillations when tracking CRH. Additionally, the model tracks heart rate, blood pressure, body temperature, and pain perception. Studied quantities function on timescales ranging from minutes to days. To understand how endotoxin impacts the body over this vast span of timescales, we examine the response to variations in LPS administration methods (single dose, repeated dose, and continuous dose) as well as the timing of the administration and the amount of endotoxin released into the system. We calibrate the model to literature

Kristen A. Windoloski
Department of Mathematics, North Carolina State University, 2311 Stinson Drive, Raleigh, NC 27607
e-mail: kawindol@ncsu.edu

Elisabeth O. Bangsgaard
Technical University of Denmark, Asmussens Allé, 2800 Lyngby, Denmark
e-mail: eoba@dtu.dk

Atanaska Dobreva
Department of Mathematics, Augusta University - 1201 Goss Lane, Augusta, GA 30912
e-mail: adobreva@augusta.edu

Johnny T. Ottesen
Centre for Mathematical Modeling - Human Health and Disease, Roskilde University, Universitetsvej 1, 4000 Roskilde, Denmark
e-mail: johnny@ruc.dk

Mette S. Olufsen
Department of Mathematics, North Carolina State University, 2311 Stinson Drive, Raleigh, NC 27607
e-mail: msolufse@ncsu.edu

data for a 2 ng/kg LPS bolus injection. Results show that LPS administration during early morning or late evening generates a more pronounced hormonal response. Most of the LPS effects are eliminated from the body 24 hours after administration, the main impact of inflammation remains in the system for 48 hours, and repeated dose simulations show that residual effects remain more than 10 days after the initial injection. We also show that if the LPS administration method or total dosage is increased, the system response is amplified, posing a greater risk of hypotension and pyrexia.

## 1 Introduction

The body has a wealth of regulatory mechanisms controlling vital functions that operate on timescales that differ by a factor of $10^7$, ranging from milliseconds (action potentials) to years (aging). Studying the effects of diseases on these timescales can be challenging, even for a well-defined event such as the inflammatory response to a low-dose endotoxin challenge (typically achieved by administering lipopolysaccharides (LPS)). The immune system is complex, and its response to a pathogenic threat entering the body through an external or internal wound varies significantly depending on the pathogen type, the degree of infection, the host's age, sex, and ethnicity [33]. The body responds to the threat by activating local and systemic (innate) signaling cascades to remove the pathogen.

Most studies examining inflammatory signaling cascades focus on the short-term response (6-8 hours) using a combination of experimental and computational approaches examining dynamics in both animals and humans. In both species, inflammation can be stimulated by low-dose LPS administration. The effects have been studied both experimentally [20, 42, 67] and computationally [6, 29, 80, 101] as this stimulus provides an excellent controlled model of the inflammatory cascade. But detailed experimental studies mapping inflammatory signaling pathways have found significant differences between animals and humans [30, 85, 108]. In addition to the immune response, pathogens impact dynamic signaling within the endocrine hypothalamic-pituitary-adrenal (HPA) axis, vascular systems, temperature regulation, and pain perception threshold [33], which display hourly, daily, monthly, and yearly variations. Long-term variations (monthly and yearly) are significant for chronic inflammation, but controlling the experimental environment is challenging. To address this challenge, we focused on developing a unified mathematical model examining the hourly and daily whole-body response to LPS, accounting for ultradian and circadian variation.

The immune, hormonal, and cardiovascular systems have historically been studied individually and often at different timescales. Mathematical modeling of the inflammatory cascade has been investigated on the timescale of hours using either models that lump inflammation components into broad categories (such as general pro-inflammatory and anti-inflammatory states) [26, 31, 53, 83] or more detailed models including specific immune response cells or cytokines [10, 18, 76].

Cardiovascular dynamics are typically studied over seconds or minutes to predict flow to a specific organ [21] or examine the control of blood flow in response to a challenge, such as the Valsalva maneuver [82]. While these models provide excellent predictions of hemodynamics, they do not address how these predictions vary daily, weekly, or monthly. Moreover, cardiovascular dynamics studies typically exclude influences from other systems, even though it is well known that the immune and hormonal systems impact dynamics. For example, the formation of atherosclerotic lesions involves an immune response [37, 16], and inflammation developing into sepsis depends on vagal responses [15, 104]. Additionally, elevated cortisol levels during stress result in increases in heart rate and anti-inflammatory reactions [60], and the transition to an advanced disease state is often accompanied by noticeable physiologic immune responses. Furthermore, morbidity is transformed into comorbidities often due to couplings by compromised immune or endocrine systems [34]

Previous studies have investigated the coupling of stress to inflammation [6, 66], and inflammation to cardiovascular dynamics, temperature, and pain perception [29, 94]. However, mathematical model coupling interactions between inflammation, stress, cardiovascular, pain, and thermal dynamics have yet to be investigated. Therefore, our study is the first to develop a mathematical model, henceforth denoted as the unified model (depicted in Figure 2) mapping the LPS response to the immune cascade, the HPA axis, and the cardiovascular system as well as temperature and pain dynamics on a timescale of hours to days. The unified model has several components: (1) an inflammation model that tracks concentrations of tumor necrosis factor-alpha (TNF-$\alpha$), interleukin 6 (IL-6), and interleukin 10 (IL-10) as well as resting and activated monocytes released in response to LPS; (2) an endocrine HPA axis model tracking concentrations of corticotropin-releasing hormone (CRH), adrenocorticotropic hormone (ACTH), and cortisol; (3) a cardiovascular model predicting heart rate, nitric oxide concentrations, vascular resistance, blood flow, and blood pressure using a circulation model integrated with a simple autonomous nerve system model; (4) a temperature model; and (5) a pain perception model. These systems operate on multiple timescales but are modeled on the timescale of hours.

Figures 2 and 2 show the coupling between the models including the stress hormone, cortisol, having a stimulating effect on heart rate, autonomous nerve system signaling affecting CRH and cytokine production, and inflammation affecting heart rate, temperature, and the HPA axis hormones. To test the validity of our unified model, we fit dynamics to data from Janum et al. (2016) and Clodi et al. (2008).

**Physiology** is the science of functions and mechanisms of the living. It dates back to Hippocrates in the late 5th century BC. The word comes from the Ancient Greek word φύσιη(phúsis), meaning ”nature” and λογία (logía), meaning ”study of”. The modern term, coined by pioneers including Jean Ferne (1497-1558), William Harvey (1578-1657), Claude Bernard (1813–1878), and August Krogh (1874-1949) refers to a model-based point of view.

The human body consists of a wealth of interconnecting mechanisms and subsystems, but these may be considered isolated or only weakly connected for many purposes. Thus, researchers have studied the circulation of blood, endocrine systems, the immune system, and other mechanisms independently. However, the development of pathologies and comorbidities often needs such subsystems to be bridged. In the current paper, we illustrate and discuss this bridging in the case of infection with an engineered coli bacterium denoted LPS. While the subsystems are well-known for specific purposes, the coupling of these quantitatively is only vaguely known due to the lack of direct in vivo measurement methodologies.

The only way the couplings of the subsystems can be described quantitatively is through mathematical modeling coupled with data for the subsystems, a strategy denoted the mathematical microscope [72]. If these couplings are of significant strength, non-linearities and multiple times scale become crucial. These complications may introduce unexpected phenomena. The current paper presents the state of such rising research.

## 2 Methods

To understand how daily (ultradian and circadian) rhythms and stress impact inflammation and how inflammation impacts cardiovascular dynamics, we develop a unified model (shown in Figures 2 and 2) integrating and adapting Dobreva et al.'s inflammatory-cardiovascular-temperature-pain model [29] and Bangsgaard et al.'s inflammatory-HPA axis model [6]. The unified model is calibrated to data from human studies administrating a low dose of LPS. Below, we describe the data used for model calibration and each model component. Model parameter values, units, and initial conditions for all state variables are listed in the Appendix, Table 1.

**Inflammation** The immune response of the human body is often divided into two sub-systems: the ***innate immune response*** and the ***adaptive immune response***. Innate immunity is the first line of defense that meets a foreign substance. It has no immunological memory and is a mechanism that naturally occurs in the body. The adaptive immune response is the next line of defense if innate immunity fails, providing a complete immune response. When the first exposure to a particular virus or pathogen occurs, the adaptive immune response creates an immunological memory to fight future threats. The body's innate immune response is the focus of our inflammation submodel presented in this paper.

The innate immune system's first defense is generally a physical barrier to block an intruder from entering the body. This could be the skin, epithelial barriers, or chemical mechanisms such as blood clotting (coagulation). The innate immune response enacts a cascade of events when the foreign substance enters the body's tissues. Phagocytic cells, whose central role is to engulf and eliminate foreign substances, destroy most of the intruders. A specific type of phagocytic cell is the macrophage, which is derived from blood monocytes. During the innate immune response, macrophages release cytokines (signaling proteins), among other chemical mediators, to not only help recruit additional immune cells to the site of infection to eliminate the threat but to also help the body repair any damage caused by the foreign substance. [68]

The immune system affects every system in the body. The response to an infection is followed by an increase in temperature (fever) and pain perception. However, other mechanisms may also be affected, including heart rate and blood pressure increases. Another essential system affected is the endocrine system, which regulates stress. Many studies have examined these systems, but little work is done to understand how they work together. Studying how the different subsystems interact can be done experimentally or, as we do here, using a complex mathematical model.

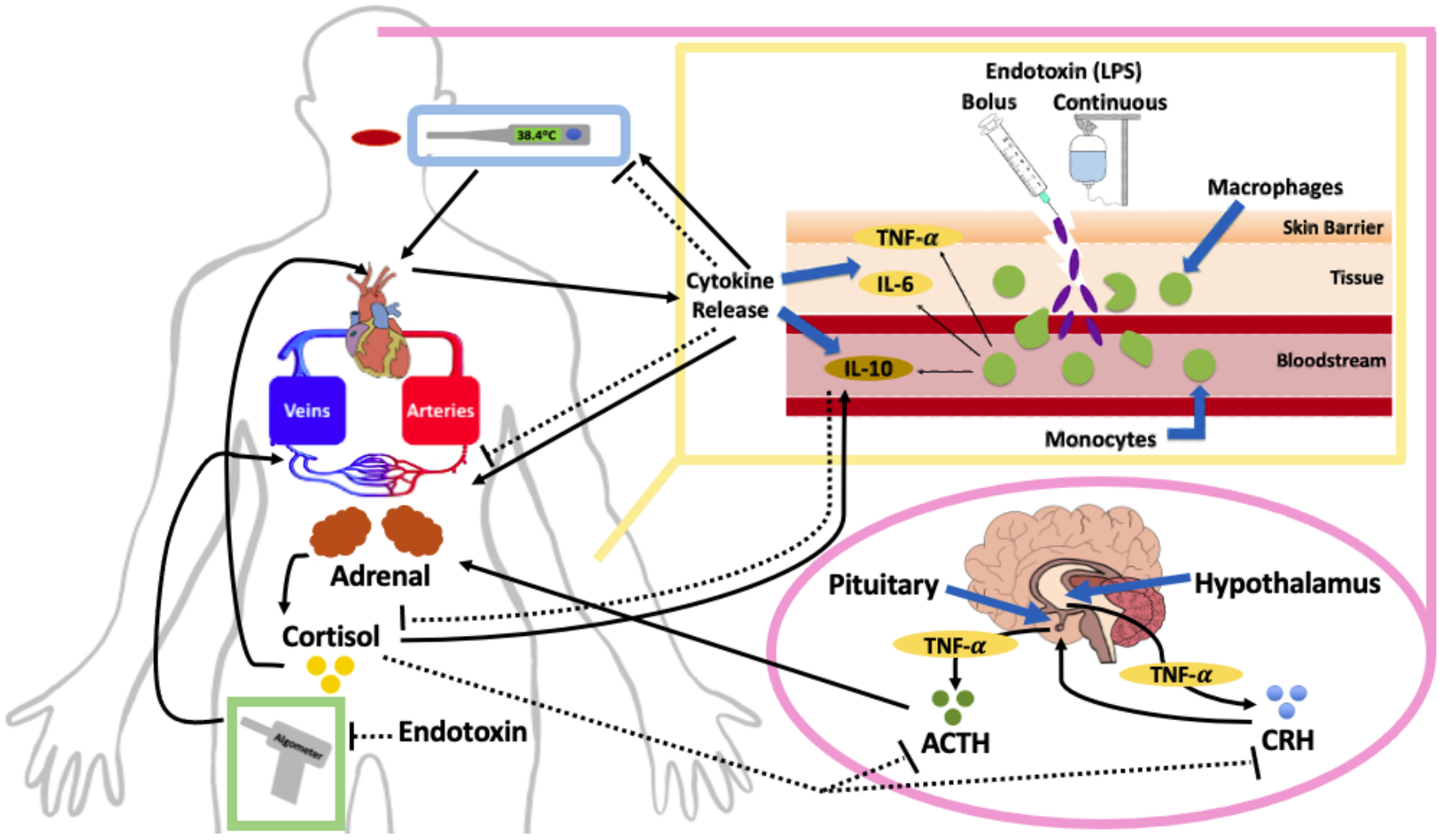

**Fig. 1** Diagram showing interactions between the immune system (yellow), the HPA axis (pink), and the cardiovascular system (red and dark blue) during an endotoxin challenge. A bolus or continuous LPS dose is administered, prompting the activation of immune cells and the secretion of pro- and anti-inflammatory cytokines. LPS administration instigates the release of pro-inflammatory cytokines, stimulating the HPA axis to produce CRH, ACTH, and cortisol. Cortisol exhibits negative feedback on CRH and ACTH and positive feedback on anti-inflammatory cytokine IL-10 and heart rate. The cytokine production also causes an increase in body temperature (light blue), which upregulates heart rate. The heart rate exhibits positive feedback on the pro-inflammatory cytokine TNF-$\alpha$. Additionally, LPS administration inhibits the pain perception threshold (green), which upregulates peripheral vascular resistance. The latter affects nitric oxide production, which is upregulated by TNF-$\alpha$ and downregulated by IL-10. The stimulation between elements is denoted by solid black lines and inhibition by dotted lines.

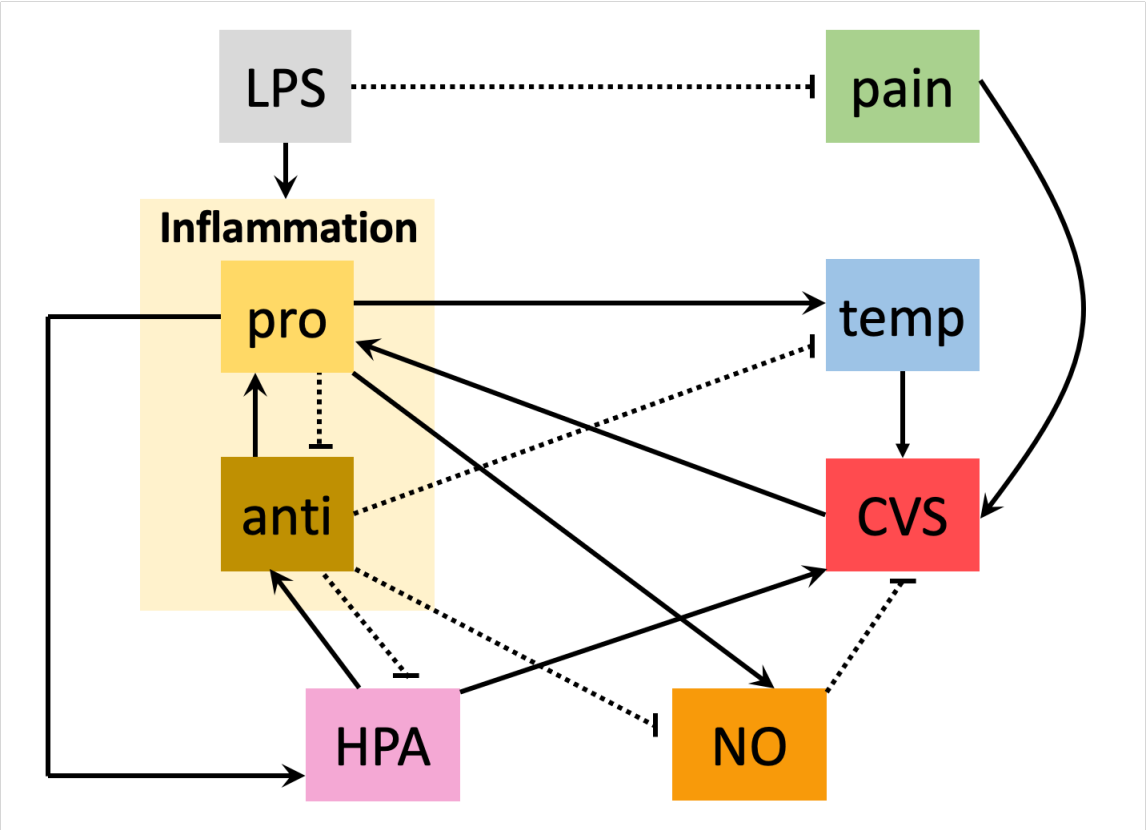

**Fig. 2** Model schematic. The inflammation model (yellow) tracks resting and activated monocytes as well as pro- (TNF-$\alpha$ and IL-6) and anti- (IL-10) inflammatory cytokines. The HPA axis model (pink) tracks CRC, ACTH, and cortisol. The cardiovascular model (red) tracks blood pressure and flow. This model is coupled with an autonomic control model predicting heart rate, nitric oxide (orange), and peripheral vascular resistance. In addition, we include a pain perception (green) and temperature (blue) model. Stimulation (upregulation) is marked by solid lines and inhibition (downregulation) by dotted lines.

## 2.1 Data

Data are extracted from the studies by Clodi et al. [20] and Janum et al. [42]. We report data of importance for constructing the unified model, but these manuscripts also include data not used in our study. In brief, Clodi et al. [20] examines the immune response to oxytocin, while Janum et al. [42] investigates the connection between pain perception and the immune response with and without nicotine. For details on these studies, we refer to their original manuscripts.

These two experimental studies measure the human response to a low dose of endotoxin (2 ng/kg). Results are reported at least hourly for 6 hours after LPS administration. The studies were approved by the respective Institutional Review Boards, and all subjects consented to participate. The study by Clodi et al. [20] was approved by the Institutional Review Board at the University of Vienna, Austria, and the study by Janum et al. [42] by the Regional Committee on Health Research Ethics and the Regional Data Monitoring Board at the University of Copenhagen, Denmark. The study by Clodi et al. [20] analyzes data from 10 male participants aged 20 to 40 years, and Janum et al. [42] analyzes data from 20 male participants aged 18 to 35 years. Participants were screened for abnormal health conditions for both studies and excluded if on any medication.

Both studies include measurements of pro-inflammatory cytokines TNF-$\alpha$ and IL-6, but Janum et al. [42] also report measurements for IL-10. Despite administering the same dose, the two studies' average inflammatory response varies significantly. The IL-6 data from Janum et al. and Clodi et al. peaks at approximately two and three hours, respectively. While these differences can be attributed to variations in the inflammatory response between individuals or be a result of differences in blood sample assaying [3], we also note that these blood samples were only collected hourly. Therefore, peak IL-6 concentrations could have occurred between the hourly blood samples, which would result in similar peak times.

Both studies report measurements of temperature, and Janum et al. [42] also reports pain perception threshold, heart rate, and blood pressure data. The study by Clodi et al. [20] mentions heart rate measurements, but values are not reported. However, Clodi et al. does report ACTH and cortisol concentrations. Figure 3 shows the data used for model calibration. While it is ideal to use data from one single study when constructing a mathematical model, there is not an endotoxin study (to our knowledge) that reports all the desired quantities for each of our submodels.

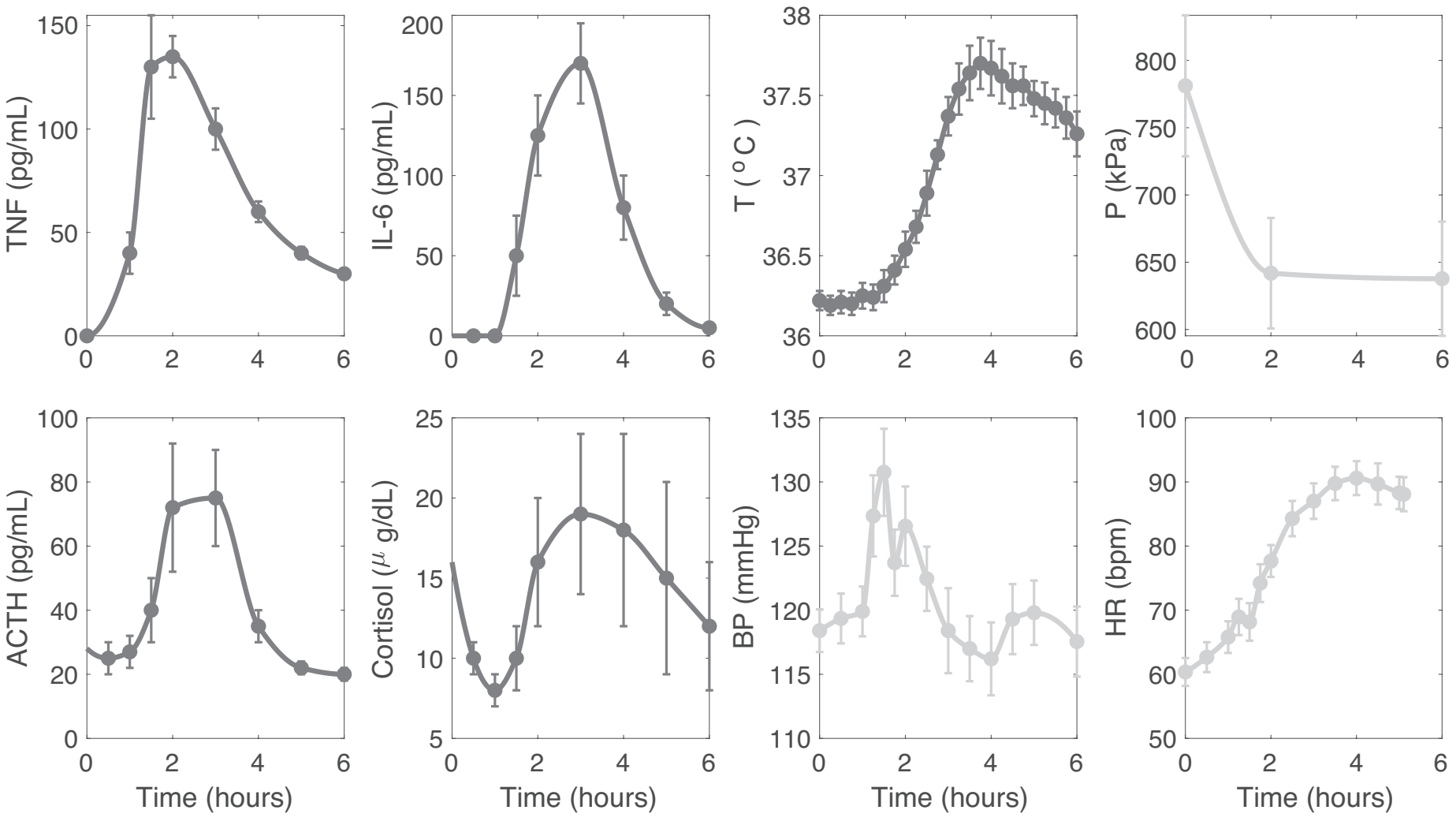

**Fig. 3** Data from [20] and [42] (mean ± SE). Data from [20] are shown in dark gray and data from [42] are shown in light gray.

**Mathematical modeling** One way to study dynamics in a complex system is via mathematical modeling, which uses mathematical equations to describe how each state variable changes in time and how the variables affect each other. Examples of state variables in this paper include cytokines in the immune system, stress hormones, body temperature, and cardiovascular quantities. Each equation formed is a differential equation, which relates a rate of change of the state variable to the value of other state variables (and possibly its effect on itself). Every state variable will have an equation, forming a system of differential equations, which can be solved using computational methods. Each equation will have parameters describing rates (how fast the variable changes), coupling strength (how much does one variable affect another), or half-saturation (how much of a given quantity is needed to get half of the maximum effect). Our paper also has parameters denoting the immunostimulant dose size and administration time. This study uses computer simulations to assess how changes in certain parameters relate to dose size and timing impact dynamics.

## 2.2 Inflammation model

**Endotoxin administration.** This study examines the response to low bolus and continuous LPS ($E$, ng/kg) stimuli. Similar to our previous study [29], and the studies by Day et al. [26] and Kadelka et al. [45], we assume that LPS decays exponentially at the rate $k_E$ (hr$^{-1}$) once administered. Therefore, we let

$$\frac{dE}{dt} = \begin{cases} k_d - k_E E, & \text{for} \quad t \le t_{cd} \\ -k_E E, & \text{for} \quad t > t_{cd}. \end{cases} \tag{1}$$

For most simulations, the total dose administered is $E_T = 2$ ng/kg. If the stimulus is given as a bolus injection, $E(0) = E_T$ and $t_{cd} = 0$ (hr). For a continuous infusion, $E(0) = 0$ and $t_{cd}$ denotes the time over which the total dose $E_T$ is administered. Here, $k_d = E_T/t_{cd}$ (ng/kg·hr) is the amount of LPS administered each hour. Note, $k_E$ has the same value independent of how the dose is administered (as bolus injection or continuous infusion).

**Inflammation cascade.** The response to endotoxin stimulation is a cascade of events, including monocyte (number of cells - abbreviated $noc$) activation and pro- and anti-inflammatory cytokine production. This study tracks TNF-$\alpha$ ($TNF$), IL-6 ($IL6$), and IL-10 ($IL10$) concentrations (pg/mL) over time (hr). Equations are set up following the interactions shown in Figure 4 using the same approach as our previous studies [6, 10, 29].

Endotoxin administration activates monocytes ($M_A$, $noc$) at a rate $k_M$ (hr$^{-1}$) recruited from the resting monocyte ($M_R$, $noc$) population [9]. The monocyte recruitment is upregulated by TNF-$\alpha$ at a rate $k_{MTNF}$ (hr$^{-1}$) and down-regulated by IL-10 [52, 63]. The resting monocytes are regenerated at a rate $k_{MR}$ (hr$^{-1}$) until the baseline level of resting monocytes, $M_\infty$ ($noc$), is reached. The activated monocytes decay at a rate $k_{MA}$ (hr$^{-1}$) without stimulation. Therefore, $M_R$ and $M_A$ dynamics are given by

$$\frac{dM_R}{dt} = k_{MR}M_R\left(1 - \frac{M_R}{M_\infty}\right) - H_M^U(E)\left(k_M + k_{MTNF}H_M^U(TNF)\right)H_M^D(IL10)M_R$$
$$\frac{dM_A}{dt} = H_M^U(E)\left(k_M + k_{MTNF}H_M^U(TNF)\right)H_M^D(IL10)M_R - k_{MA}M_A.$$

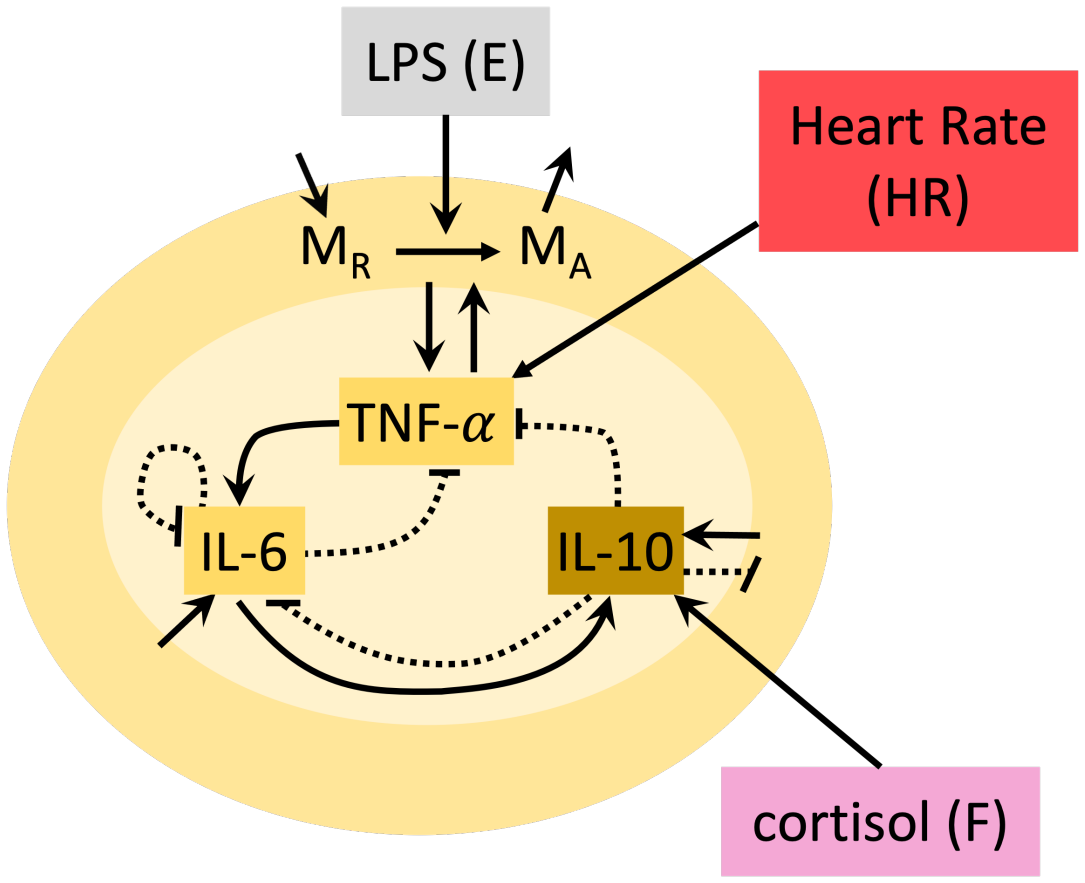

**Fig. 4** Inflammatory model. LPS $(E)$ stimulates monocyte activation $(M_A)$, which upregulates pro- (TNF-$\alpha$ and IL-6) and anti- (IL-10) inflammatory cytokines. The cytokines regulate one another through positive and negative feedback. Note that IL-6 is pro- and anti-inflammatory, downregulating itself and TNF-$\alpha$ but upregegulate IL-10. Additionally, the inflammatory response is regulated by the HPA axis and heart rate. Solid lines denote stimulation (upregulation) and dotted lines inhibition (downregulation).

In the above equations and throughout this study, upregulation (stimulation) is denoted by $H_Y^U(X)$ and downregulation (inhibition) by $H_Y^D(X)$. These stimuli are modeled by Hill functions of form

$$H_Y^U(X) = \frac{X^h}{X^h + \eta_{YX}^h}$$
$$H_Y^D(X) = \frac{\eta_{YX}^h}{X^h + \eta_{YX}^h},$$

where $\eta_{YX}$ denotes the half-saturation value of variable $X$ and $h$ the Hill exponent determining the steepness of the effect on $Y$.

Activated monocytes stimulate TNF-$\alpha$ production at a rate $k_{TNFM}$ (pg/mL·hr·noc) [12]. Both IL-6 and IL-10 downregulate TNF-$\alpha$ production, and TNF-$\alpha$ naturally decays to baseline level $w_{TNF}$ (pg/mL) at rate $k_{TNF}$ (hr$^{-1}$) [19, 93, 97]. Activated monocytes stimulate IL-6 production at a rate $k_{6M}$ (pg/mL·hr·noc), and TNF-$\alpha$ stimulates IL-6 production at a rate $k_{6TNF}$ (pg/mL·hr·noc) [97]. Moreover, IL-6 exhibits anti-inflammatory properties, including downregulation of itself, and IL-6 naturally decays to baseline level $w_6$ (pg/mL) at a rate $k_6$ (hr$^{-1}$) [99].

In addition to the inflammatory feedback, the cardiovascular system modulates pro-inflammatory cytokines. Typically, an inflammatory event increases heart rate ($HR$, bpm) above its baseline value ($HR_b$, bpm), which in turn impacts TNF-$\alpha$ production at a rate $k_{TNFH}$ (bpm$^{-1}$) [44, 79]. When the heart rate is at the baseline level, it does not impact TNF-$\alpha$ production. Therefore, pro-inflammatory cytokines TNF-$\alpha$ and IL-6 are determined by

$$\begin{aligned}\frac{dTNF}{dt} &= k_{TNFM}H^D_{TNF}(IL6)H^D_{TNF}(IL10)\left(1 + k_{TNFHR}(HR - HR_b)\right)M_A \\ &\quad - k_{TNF}(TNF - w_{TNF}) \\ \frac{dIL6}{dt} &= \left(k_{6M} + k_{6TNF}H^U_{IL6}(TNF)\right)H^D_{IL6}(IL6)H^D_{IL6}(IL10)M_A - k_6(IL6 - w_6).\end{aligned}$$

The anti-inflammatory cytokine IL-10 is stimulated by activated monocytes at a rate of $k_{10M}$ (pg/mL·hr·noc) and upregulated by IL-6 at the rate $k_{106}$ (pg/mL·hr·noc) [13, 43]. IL-10 decays to baseline level $w_{10}$ (pg/mL) at the rate $k_{10}$ (hr$^{-1}$). In addition, cortisol ($F$, $\mu$g/dL) has anti-inflammatory properties [50, 59], and its influence on IL-10 dynamics is modeled by upregulation of IL-10 production at the rate $k_{10F}$ (pg/mL·hr·noc). Thus, the equation for IL-10 is given by

$$\frac{dIL10}{dt} = \left(k_{10M} + k_{106}H^U_{IL10}(IL6) + k_{10F}H^U_{IL10}(F)\right)M_A - k_{10}(IL10 - w_{10}).$$

## 2.3 HPA axis model

We briefly summarize the HPA axis model equations below, but a detailed description can be found in [6, 73] with further background in [2, 5, 38, 66, 39]. The HPA axis model is represented by three differential equations tracking concentrations of CRH ($C$, pg/mL), ACTH ($A$, pg/mL), and cortisol ($F$, $\mu$g/dL) as

$$\begin{aligned}\frac{dC}{dt} &= k_{CR}R(t)H^D_C(F)C + k_{CTNF}TNF - k_C(C - C_b) \\ \frac{dA}{dt} &= k_{AC}H^D_A(F)C + k_{ATNF}H^U_A(TNF) - k_A A \\ \frac{dF}{dt} &= k_{FA}H^D_F(IL10)A^2 - k_F F.\end{aligned}$$

The release of CRH by the hypothalamus is influenced by the circadian rhythm, denoted by the time-dependent function $R(t)$ (discussed in further detail below), at the rate $k_{CR}$ (hr$^{-1}$) and is also stimulated by the presence of TNF-$\alpha$ through the rate $k_{CTNF}$ ($\text{hr}^{-1}$) [7, 104]. Cortisol downregulates CRH, and CRH decays to a baseline level $C_b$ (pg/mL) at rate $k_C$ ($\text{hr}^{-1}$). Stimulation of the pituitary gland by CRH leads to production of ACTH at the rate $k_{AC}$ ($\text{hr}^{-1}$) while cortisol downregulates this process [102, 105]. ACTH production is also upregulated by TNF-$\alpha$ at the rate $k_{ATNF}$ (pg/mL·hr) [7], and ACTH levels decay at a rate of $k_A$ ($\text{hr}^{-1}$). Finally, cortisol is downregulated by the anti-inflammatory cytokine IL-10 and stimulated from the adrenal glands by ACTH at rate $k_{FA}$ ($\mu$g·mL/pg·dL·hr) [41]. In the absence of stimulation, cortisol levels decay at the rate $k_F$ ($\text{hr}^{-1}$) [6]. A schematic of the HPA axis model is presented in Figure 5.

The time-dependent function $R(t)$ (non-dimensional, n.d.) denotes an enforced exogenous circadian rhythm (the body's 24-hour cycle) that impacts mental, physical, and behavioral processes in humans[1, 71]. It is modeled by

$$R(t) = \left( \frac{t_m^k}{t_m^k + \alpha^k} \cdot \frac{(T_{24} - t_m)^\ell}{(T_{24} - t_m)^\ell + \beta^\ell} + \varepsilon \right) N_c$$

as the product of an upregulation Hill function with a half-saturation value $\alpha$ (min) and Hill exponent $k$ (n.d.) and a downregulation Hill function with half-saturation value $\beta$ (min) and Hill exponent $\ell$ (n.d.). In the Hill functions, $t_m$ (min) denotes the time in minutes during the 24 hour cycle shifted by the value $\delta$ (min), i.e. $t_m = (60t - \delta)$ modulo $T_{24}$, with $T_{24} = 60 \cdot 24 = 1440$ min as the cycle length (24 hours converted to minutes). Finally, $\varepsilon$ (n.d.) denotes the base value of the circadian rhythm, and $N_c$ (n.d.) is a scaling factor.

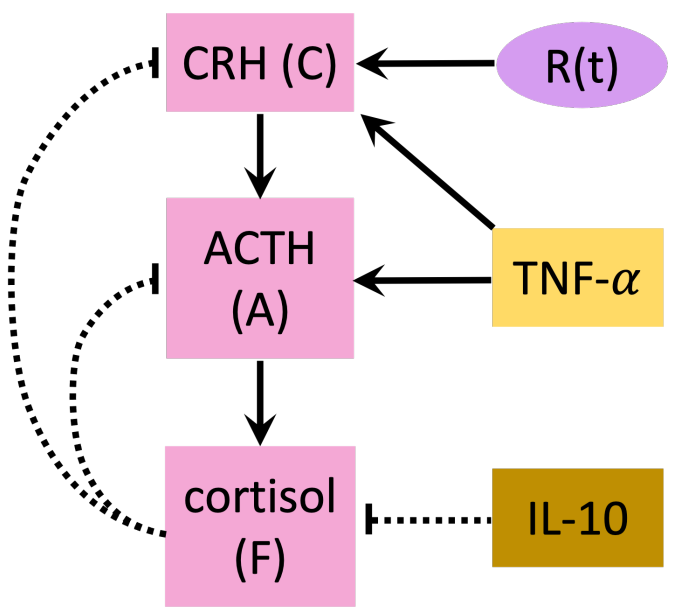

**Fig. 5** HPA axis model. CRH ($C$) production is stimulated by the body's natural circadian rhythm (modeled by the function $R(t)$) and TNF-$\alpha$. CRH and TNF-$\alpha$ also stimulate ACTH ($A$) production, which in turn stimulates cortisol ($F$) production. Cortisol is downregulated by IL-10 and exhibits negative feedback on both CRH and ACTH. Stimulation (upregulation) is denoted by solid lines and inhibition (downregulation) by dotted lines.

## 2.4 Cardiovascular model

Given the timescale of hours, the cardiovascular system and its control (shown in Figure 6) are modeled as non-pulsatile [107]. The transport model tracks changes in systemic volume ($V$, mL), pressure ($p$, mmHg), and flow ($q$, mL/hr) in the arteries and veins. Feedback from this and the model's control tracks changes in heart rate ($HR$, bpm) and peripheral vascular resistance ($R_S$, mmHg·hr/mL). The cardiovascular model consists of four compartments: the large arteries ($la$), small arteries ($sa$), small veins ($sv$), and large veins ($lv$). We make a note that while we model the cardiovascular section where $t$ is in hours, dimensional analysis can be used to model flow and, subsequently, resistance where $t$ is in seconds and heart rate where $t$ is in minutes, which is how these quantities are usually computed. The correct analysis would produce the same results as seen here in our study.

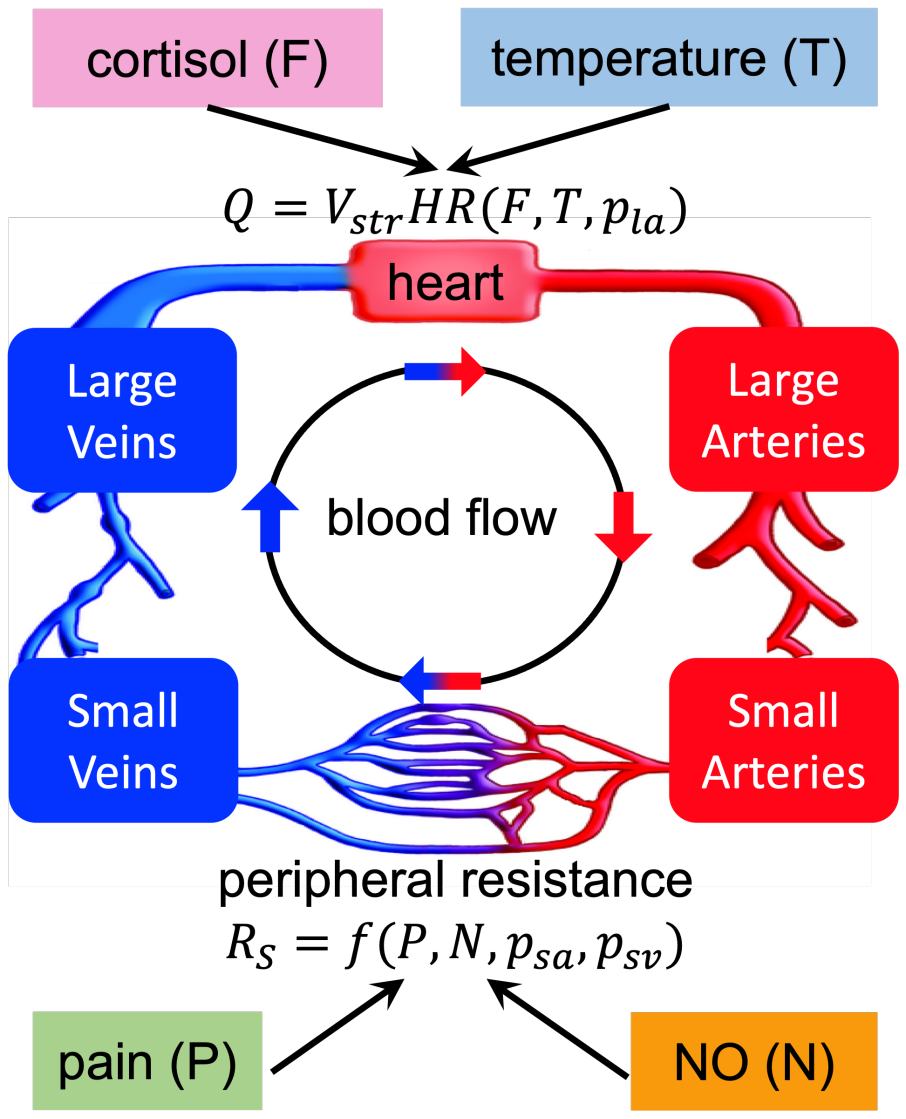

**Fig. 6** The cardiovascular model. It predicts pressure $p$ (mmHg), flow $q$ (mL/hr), and volume $V$ (mL) in four compliant compartments representing the systemic large arteries ($la$), small arteries ($sa$), large veins ($lv$), and small veins ($sv$). Flow through the heart $Q$ (mL/hr) is predicted as a function of heart rate ($HR$, bpm) and stroke volume ($V_{str}, mL/beat$). The control model regulates heart rate as a function of blood pressure, cortisol ($F$, $\mu$g/dL), and body temperature ($T$, °C). Peripheral vascular resistance ($R_S$, mmHg·hr/mL) is regulated in response to pain perception ($P$, kPa) and nitric oxide (N, n.d.).

**Cardiovascular transport.** The system dynamics are modeled using a hydrodynamic analog to a resistor-capacitor (RC) circuit. Voltage is analogous to pressure, current to flow, capacitance to compliance (or elastance, $E$ (mL/mmHg), which is the reciprocal of compliance), while resistance $R_S$ refers to resistance in both formulations. A system of differential equations is obtained by ensuring conservation of volume, given by

$$\frac{dV_{la}}{dt} = Q - q_a$$
$$\frac{dV_{sa}}{dt} = q_a - q_s$$
$$\frac{dV_{sv}}{dt} = q_s - q_v$$
$$\frac{dV_{lv}}{dt} = q_v - Q,$$

where $V_i$ (mL) denotes the total volume of compartment $i \in \{la, sa, sv, lv\}$, $Q$ (mL/hr) is the cardiac output (flow) through the heart, and $q_j$ (mL/hr), $j \in \{a, s, v\}$ denotes the flow between the compartments. Flow is related to pressure via Ohm's law, given by

$$q_j = \frac{p_{out} - p_{in}}{R_j}, \tag{2}$$

where $R_j$ (mmHg·hr/mL) is the resistance between the $j$'th compartments. Pressure is related to volume via the pressure-volume equation

$$V_i - V_{un,i} = E_i(p_i - p_{un,i}), \tag{3}$$

where $E_i$ is the compartment elastance, and $V_{un,i}$ and $p_{un,i}$ denote the unstressed volume and pressure, respectively. Cardiac output is determined by $Q = HR \cdot V_{str}$, where $HR$ (bpm) is the heart rate and $V_{str}$ (mL/beat) is the stroke volume. Following the derivation in our previous study [107], stroke volume is given by

$$V_{str} = V_{ED} - V_{ES} = -\left(\frac{p_{la}}{E_{Max}} - \frac{p_{lv}}{E_{Min}}\right), \tag{4}$$

where $V_{ED}$ and $V_{ES}$ denote the end-diastolic and end-systolic volume (mL), $p_{la}$ and $p_{lv}$ are the systemic arterial and systemic venous pressures (mmHg), and $E_{Max}$ and $E_{Min}$ are the maximum (end-diastolic) and minimum (end-systolic) elastance (mmHg/mL).

**Cardiovascular control.** Since we analyze dynamics over hours rather than seconds, we ignore the fast response (within seconds) to sudden changes in blood pressure. The longer-term effects (on the order of minutes to hours) include inflammatory, cortisol, and temperature modulation of pseudo-steady levels of noradrenaline and acetylcholine regulating heart rate and vascular resistance via the sympathetic and parasympathetic systems. In addition, vascular resistance is modulated by nitric oxide (NO), an effective vasodilator.

*Heart rate* is upregulated by changes in temperature ($T$, °C), blood pressure ($p_{la}$, mmHg), and cortisol ($F$). The rostral raphe region of the medulla oblongata has temperature-regulating sympathetic neurons located close to the cardiac-related sympathetic neurons. Therefore, an increase in body temperature triggered by inflammation upregulates heart rate via sympathetic activation and parasympathetic inhibition [14, 25, 69]. Similarly, cortisol upregulates heart rate via an increase in noradrenaline from the adrenal glands [40, 56, 92], while blood pressure impacts heart rate via two mechanisms [24, 47, 70, 110]. If the blood pressure is higher than its baseline, heart rate increases [70], but if blood pressure falls below the resting value, commonly observed in patients with active inflammation [74], heart rate is also upregulated. To include these effects, the change in heart rate is predicted as

$$\frac{dHR}{dt} = \frac{1}{\tau_H}\Big(k_H(HR_M - HR_b)H^U_{HR}(T - T_b)H^U_{HR}(F)f(p_{la}) - (HR - HR_b)\Big), \tag{5}$$

where $\tau_H$ (hr$^{-1}$) and $k_H$ (n.d.) are rate constants, $HR_M$ (bpm) is the maximal heart rate, $HR_b$ (bpm) is the baseline heart rate, $T$ is temperature (°C), $T_b$ (°C) is the baseline temperature, and the control responding to changes in blood pressure $f(p_{la})$ is given by

$$f(p_{la}) = \begin{cases} H^U_H(p_{lab} - p_{la}), & p_{la} \le 100 \text{ mmHg} \\ H^D_H(p_{la} - p_{lab}), & p_{la} > 100 \text{ mmHg}, \end{cases}$$

where $p_{la}$ is the arterial blood pressure predicted by the cardiovascular model and $p_{lab}$ (mmHg) is the baseline arterial blood pressure.

*Peripheral vascular resistance*, $R_S$, is primarily regulated by pain [62, 87] and nitric oxide [57]. It increases with an increase in the rate of change of pain perception threshold $\Gamma = \frac{dP}{dt}$ (kPa/hr) with half saturation value $\eta_{RP}$ (kPa/hr) and growth rate $k_{RP}$ (mmHg/mL). Additionally, nitric oxide ($N$, n.d.) is a well-known vasodilator [57, 65, 88, 89], and so the presence of nitric oxide decreases vascular resistance at a rate of $k_{RN}$ (mmHg/mL). Elevated levels of $R_S$ return to baseline $R_{Sb}$ (mmHg·hr/mL) at rate $k_R$ (hr$^{-1}$). These effects give

$$\frac{dR_S}{dt} = k_{RP}\frac{\Gamma^2}{\Gamma^2 + \eta^2_{RP}} - k_{RN}N - k_R(R_S - R_{Sb}), \tag{6}$$

where nitric oxide ($N$) is determined by

$$\frac{dN}{dt} = k_{NM}H^U_N(TNF(t-\kappa))H^D_N(IL10(t-\kappa))M_A - k_N N,$$

accounting for the 2-4-hour delay in nitric oxide production in response to the inflammatory event. Activated monocytes stimulate nitric oxide production via inducible NO synthase [65]. It is upregulated by TNF-$\alpha$ and downregulated by IL-10 [17, 88]. The production rate is determined by the constant $k_{NM}$ ((hr·noc)$^{-1}$). The delay in NO stimulation and suppression from cytokine concentrations TNF-$\alpha$ and IL-10 is determined by $\kappa$. In the absence of stimulation, nitric oxide levels decay to baseline at the rate $k_N$ (hr$^{-1}$).

We note that in the remaining model simulation figures (Figures 7–12), vascular resistance $R_S$ is reported in seconds instead of hours. This was done to decrease the magnitude of the $R_S$ solution and was calculated by dividing the $R_S$ solution by 3600 (converting hours to seconds).

## 2.5 Temperature and pain

**Temperature.** The production of pro-inflammatory cytokines TNF-$\alpha$ and IL-6 stimulates and sustains [22, 55, 93] a fever while anti-inflammatory cytokine IL-10 [36, 75] helps to regulate body temperature. Therefore, the body temperature $T$ (°C) changes in response to endotoxin as

$$\begin{aligned}\frac{dT}{dt} &= k_{TTNF}H_T^U(TNF - w_{TNF}) + k_{T6}H_T^U(IL6 - w_6)\\ &\quad -k_{T10}H_T^U(IL10 - w_{10}) - k_T(T - T_b),\end{aligned}$$

where $k_{TTNF}$, $k_{T6}$, and $k_{T10}$ describe the rate (°C/hr) at which TNF-$\alpha$, IL-6, and IL-10 affect body temperature, respectively. Temperature returns to baseline level $T_b$ (°C) at the rate $k_T$ (hr$^{-1}$) . As mentioned above, TNF-$\alpha$ and IL-6 upregulate temperature while IL-10 downregulates temperature. Each Hill function is shifted by the baseline cytokine value so that when cytokines are at the baseline level, they do not stimulate a change in body temperature. The equation is adapted from the more complex form in [29]. Temperature dynamics are shown in Figure 2.

**Pain.** Several studies have investigated the connection between immune response and pain perception [54, 100]. In response to either exposure to endotoxin or the cytokine cascade resulting from endotoxin detection, pain receptors (nociceptors) become activated [8, 42, 106]. We predict pain perception using the formulation from Dobreva et al. [29] resulting in

$$\frac{dP}{dt} = -k_{PE}E\,P - k_P(P - P_b).$$

The presence of endotoxin ($E$, ng/kg) decreases the pain perception threshold ,$P$ (kPa), at the rate $k_{PE}$ (kg/ng·hr). As the endotoxin is eliminated from the system, $P$ returns to baseline value $P_b$ (kPa) at the recovery rate $k_P$ (hr$^{-1}$).

## 2.6 Model calibration

The inflammatory-cardiovascular model originates from [29] and the HPA axis model originates from [6]. Combining these into a unified model requires adjustments and re-calibration of the model parameters. The unified model uses modified computer code from the study by Dobreva et al. [29] augmented with a modified version of the integrated HPA model from Bangsgaard et al. [6].

We use cytokine, temperature, and hormonal data from Clodi et al. [20], and cardiovascular and pain perception data from Janum et al. [42]. The decision to use cytokine data from Clodi et al. instead of Janum et al. is due to the sequence used to couple the submodels. We first calibrated the inflammation model (uncoupled from the cardiovascular model) to the Clodi et al. data and then coupled the inflammation model to the HPA axis model. Given that the Clodi et al. study had both cytokine and hormonal data, we used the cytokine data from Clodi et al. for model calibration. This involved coupling TNF-$\alpha$ with CRH and ACTH and IL-10 with cortisol. Each coupling was done successively to select the relevant parameter values. We also had to scale TNF-$\alpha$ and IL-6 down to the appropriate concentrations reported by Clodi et al. [20]. This required scaling several TNF-$\alpha$ and IL-6 related parameters. Next, we coupled the inflammatory-HPA axis model with the cardiovascular-temperature-pain model from [29] in a similar scaffolding manner. Once the models were coupled, as shown in Figure 2, influential parameters were manually adjusted to fit the model to the data.

# 3 Results

We examined the unified model's response to endotoxin dosing type, amount, and timing. The latter is of particular interest for exploring dynamics at different timescales: the fast cardiovascular response (minutes), the intermediate inflammatory, the HPA axis response (hours), and the slow response of the circadian rhythm (days), while exploration of the model's response to variations in endotoxin amount and dosing methods are of clinical relevance [4, 26, 49, 98].

All simulations used a total dose of 2 ng/kg (the dose used by both Clodi et al. [20] and Janum et al. [42]) except the simulation analyzing the impact of the total dose. Simulations are depicted over three to four 24-hour cycles, including up to two 24-hour cycles before the LPS administration and two 24-hour cycles post-LPS. Supplemental simulations (`https://kwindoloski.wordpress.ncsu.edu/`) show results on longer and shorter time intervals.

## 3.1 Single LPS administration - base simulation

We first simulate the effect of a bolus dose of 2 ng/kg of LPS given at time $t$ = 37.5 hours (1:30 pm on day 2). Figure 7 shows the data and the model output. We depict dynamics for one full cycle ($t$ = 0 to $t$ = 24 hours) before LPS administration and two full cycles following dispensation, totaling four 24-hours cycles.

Figure 7 shows that the inflammation arising from the LPS stimulus causes a rapid increase in cytokine levels in the first hours following the injection. In agreement with previous studies [6, 18, 29], most of the inflammatory response returns to baseline after about 6 hours. The same does not apply to the HPA axis hormones CRH, ACTH, and cortisol, which are strongly affected by ultradian oscillations emerging from the circadian forcing. ACTH follows cytokine dynamics timing, but CRH and cortisol are elevated for at least 24 hours following injection, after which they normalize. Temperature and heart rate also have a fast response; they recover in about 10 hours, but nitric oxide, resistance, pain perception, and blood pressure take about three 24-hour cycles to recover fully.

In this simulation, we also see an initial blood pressure increase followed by a blood pressure decrease below baseline. The initial increase is caused by a pain perception decrease and a vasoconstriction increase. The blood pressure drop is a result of the heart rate normalizing at a much faster rate than nitric oxide. This behavior is not captured explicitly in the blood pressure data since blood pressure measurements are only taken for six hours following endotoxin administration. However, decreases in blood pressure measurements below baseline have been observed in other endotoxin challenge studies [51, 61, 89]. Thus, the mechanisms included in our model and evidence from other experimental studies lead us to believe that this behavior is plausible.

In summary, the model generates an initial increase in pro- and anti-inflammatory cytokine levels, hormone levels, body temperature, nitric oxide, blood pressure, heart rate, and vascular resistance. Elevation of IL-6 induces a slight fever, causing a drop in resistance, blood pressure, and pain perception threshold below baseline levels.

## 3.2 Timing of LPS administration

To study the effect of LPS injection timing, we selected four dispensing times aligned with critical cortisol values over a 24-hour cycle. Times selected (shown in Figure 8**(a)** are at $t$ = 2 hours (low cortisol just before the circadian increase), $t$ = 7 hours (the highest level of cortisol caused by the circadian and ultradian effects), $t$ = 11.9 hours (the ultradian oscillation valley during declining circadian activation), and $t$ = 21.8 hours (just after circadian activation where the cortisol level is low). These times correspond to 2:00 am, 7:00 am, 12:00 pm, and 10:00 pm. We denote these as *early morning*, *morning*, *noon*, and *late evening*. Similar to the calibration simulation discussed above, this simulation uses the 2 ng/kg endotoxin dose.

Figures 8(**a**) and (**b**) show that the resting and activated monocytes primarily shift the inflammatory response to the right. TNF-$\alpha$ and IL-10 exhibit slightly higher peaks when LPS is administered in the morning or at noon. However, the administration time mainly impacts HPA axis hormones. ACTH and cortisol have a significant spike right after the LPS administration. If administered in the morning, the peak is during the upslope of the circadian wave, dampening the ultradian oscillations, while administering it in the early morning or late evening increases the ultradian oscillations. However, these states return to their baseline after 24 hours. Another interesting observation is the effect of LPS administration time on CRH. When administered in the morning or at noon, LPS timing has a minimal effect on CRH. Yet, when LPS is administered in the early morning or late evening, CRH concentration spikes within 8 hours to much higher levels than seen in Figure 8. As with ACTH and cortisol, CRH returns to its baseline after 24 hours.

The cardiovascular state blood pressure exhibits a slightly larger drop if LPS is administered at noon, while vascular resistance and nitric oxide are unaffected by the injection time. The heart rate peak remains similar during the different injection times, but the return to the baseline is partially stunted when LPS is administered early or late evening. Finally, pain perception and temperature are shifted with the administration times.

### 3.3 Repeated LPS administration

A few studies have investigated the impact of repeated LPS administrations on the immune response by mathematical modeling [6, 26, 29, 86], but either not coupled to the cardiovascular dynamics or not accounting for ultradian and circadian variation. To understand these effects, we examine what happens when two 1 ng/kg LPS doses are administered repeatedly. Chosen so that the total dose remains the same as in our previous investigations, the first dose is administered at a fixed time ($t_1 = 13.5$ hours) and the second dose is given at times $t_2 = 14.5, 19.5, 25.5$, and 37.5 hours corresponding to *1 hour*, *6 hours*, *12 hours*, and *24 hours* after the first LPS dose, respectively. Simulation results, shown in Figure 9, reveal that the inflammation states have pronounced peaks when the second dose is given up to six hours following the first dose. However, the second reaction is almost suppressed if the second dose is given after cytokine levels return to baseline values.

The HPA axis displays a rapid and pronounced ACTH peak reaching 100 pg/mL. Most LPS effects on ACTH have worn off by 4-6 hours after the second dose, but some effects remain for more than 48 hours. The cortisol dynamics are similar ACTH except when the repeated administration is close to the first dose. In this case, the rapid peak is modest, and normalization happens faster. The repeated dose causes elevated ultradian oscillation in the next circadian cycle. A more pronounced response in CRH is observed if the repeated dose of LPS is given while CRH is low compared to cases where CRH is elevated. When the second LPS dose is administered 12 or 24 hours after the first dose, the repeated administration almost normalizes the ultradian oscillations in CRH and, thus, ACTH and cortisol.

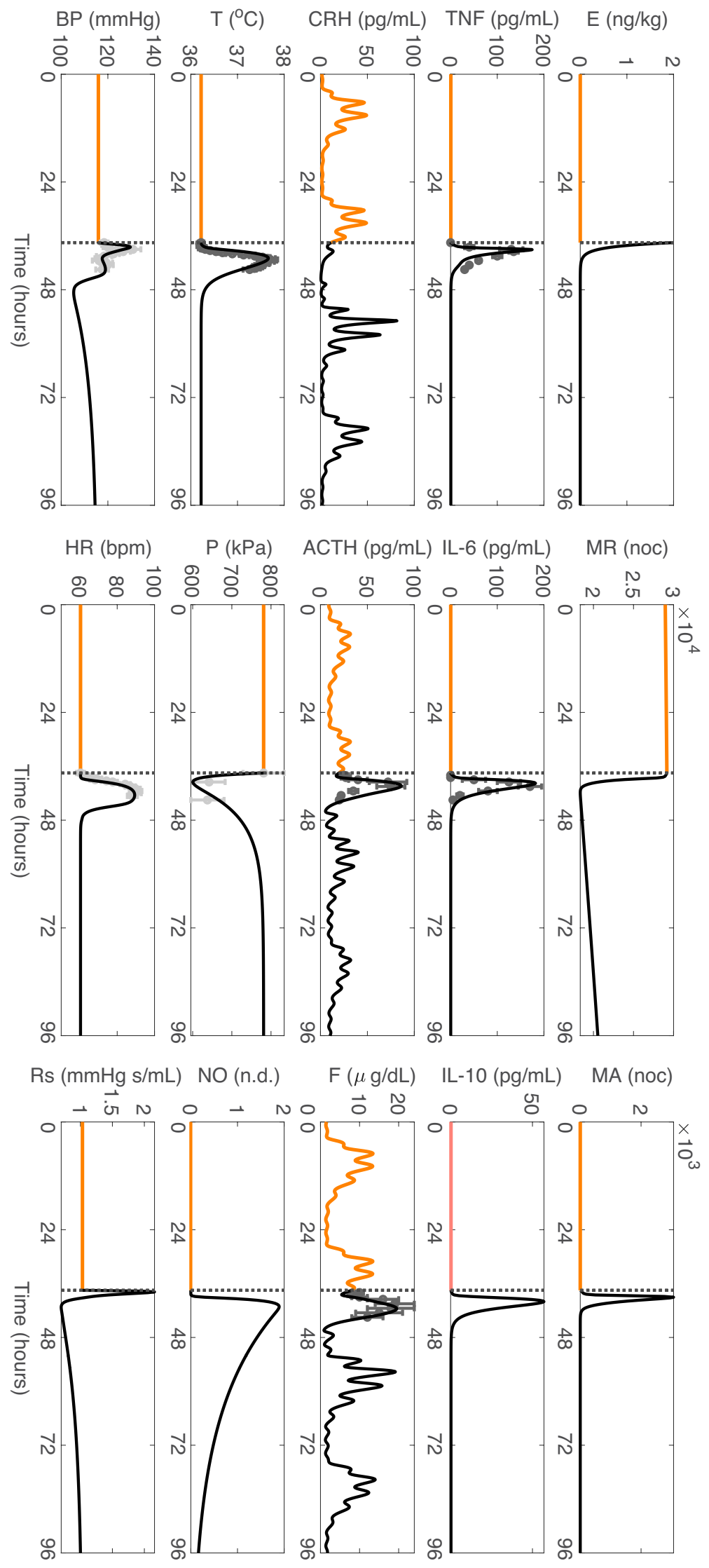

**Fig. 7** Simulation generated by solving the unified model with parameters calibrated to data. Results are depicted for four 24-hour cycles. LPS is administered at time $t = 37.5$ hours (marked by vertical dotted lines). Results at times before LPS injection are marked by solid orange lines, while solid black lines are used for results at times after LPS injection. Data from [20] are shown in dark gray and data from [42] in light gray (mean ± SE).

The cardiovascular state blood pressure drops below 100 mmHg when the second dose is given when blood pressure is high (up to 6 hours after the first dose), causing hypotension. If the second dose is given after blood pressure returns to baseline, we see a modest blood pressure increase with no significant drops below baseline. When the second dose is delivered while the heart rate is high, the heart rate increases to a slightly higher peak value. In contrast, if a second dose is administered after heart rate returns to baseline, smaller increases in heart rate occur. The response in nitric oxide follows that of cytokines, except that the decay to baseline is significantly slower (approximately 48 hours). Resistance shows a modest peak due to the repeated injections. When the second dose is given up to 6 hours after the first dose, the resistance drops by up to 50% of its baseline value, followed by a slow recovery (approximately 48 hours). If the resistance returns to baseline before the repeated dose, we see an increase in resistance without the following drop below baseline. The temperature has a similar response to the repeated dose as the cytokines. The pain perception threshold dramatically drops when the repeated injection is given close to the first one, corresponding an increased pain sensitivity. However, the pain perception threshold exhibits a smaller drop the later the repeated dose is given, and its levels return to baseline in less than 24 hours.

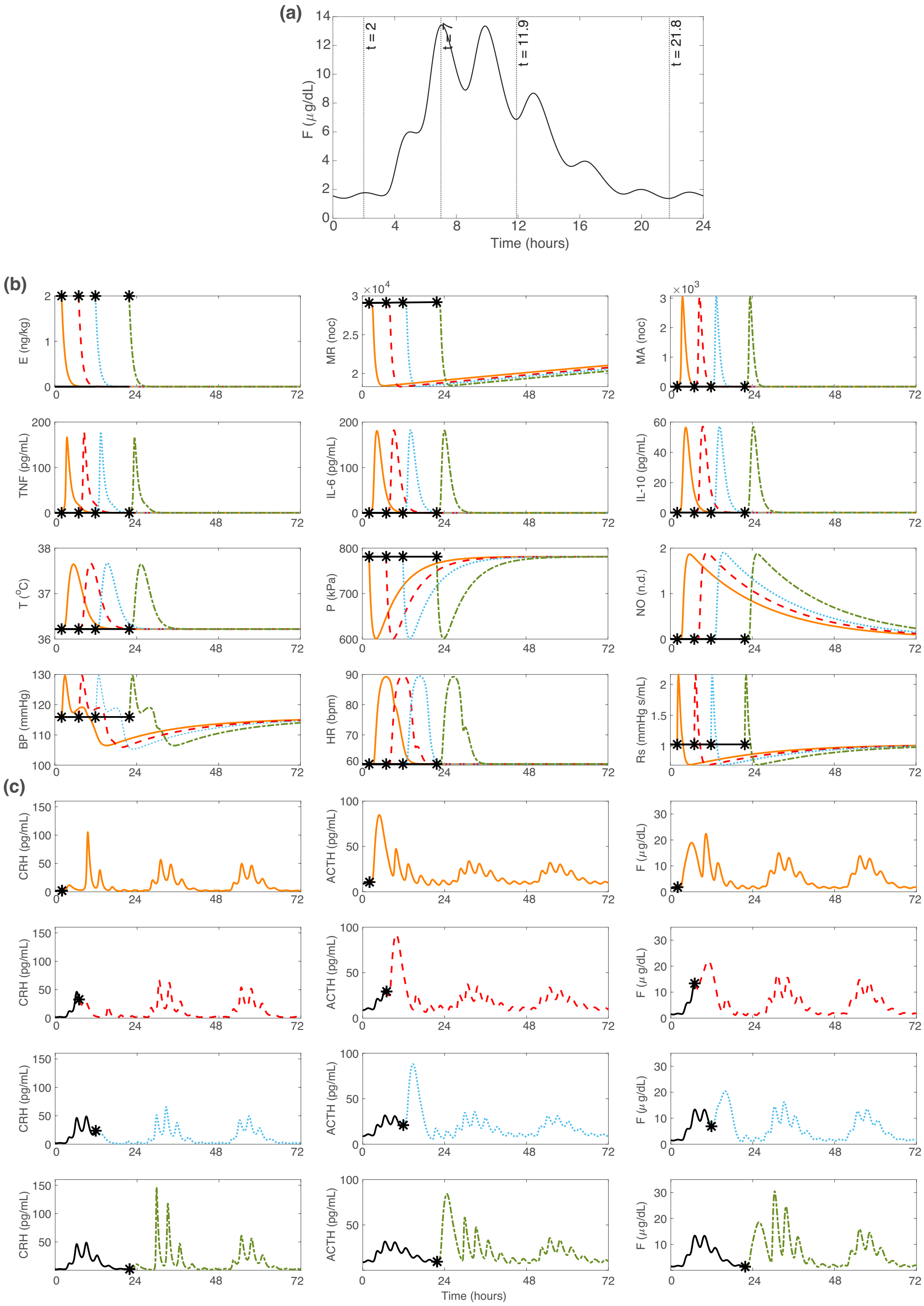

**Fig. 8** Model simulations examining the timing of LPS administration. The solid black line denotes the model simulation before LPS administration. Black asterisks mark when LPS is administered (at $t = 2, 7, 11.9$, and 21.8 hours). Post-LPS dynamics are depicted by orange solid lines ($t = 2$), red dashed lines ($t = 7$), blue dotted lines ($t = 11.9$), and green dashed-dotted lines ($t = 21.8$). (**a**) times (marked with dotted vertical lines) at which LPS is administered within a normal 24-hour cortisol cycle. (**b**) Inflammatory and cardiovascular simulations. (**c**) HPA axis simulations.

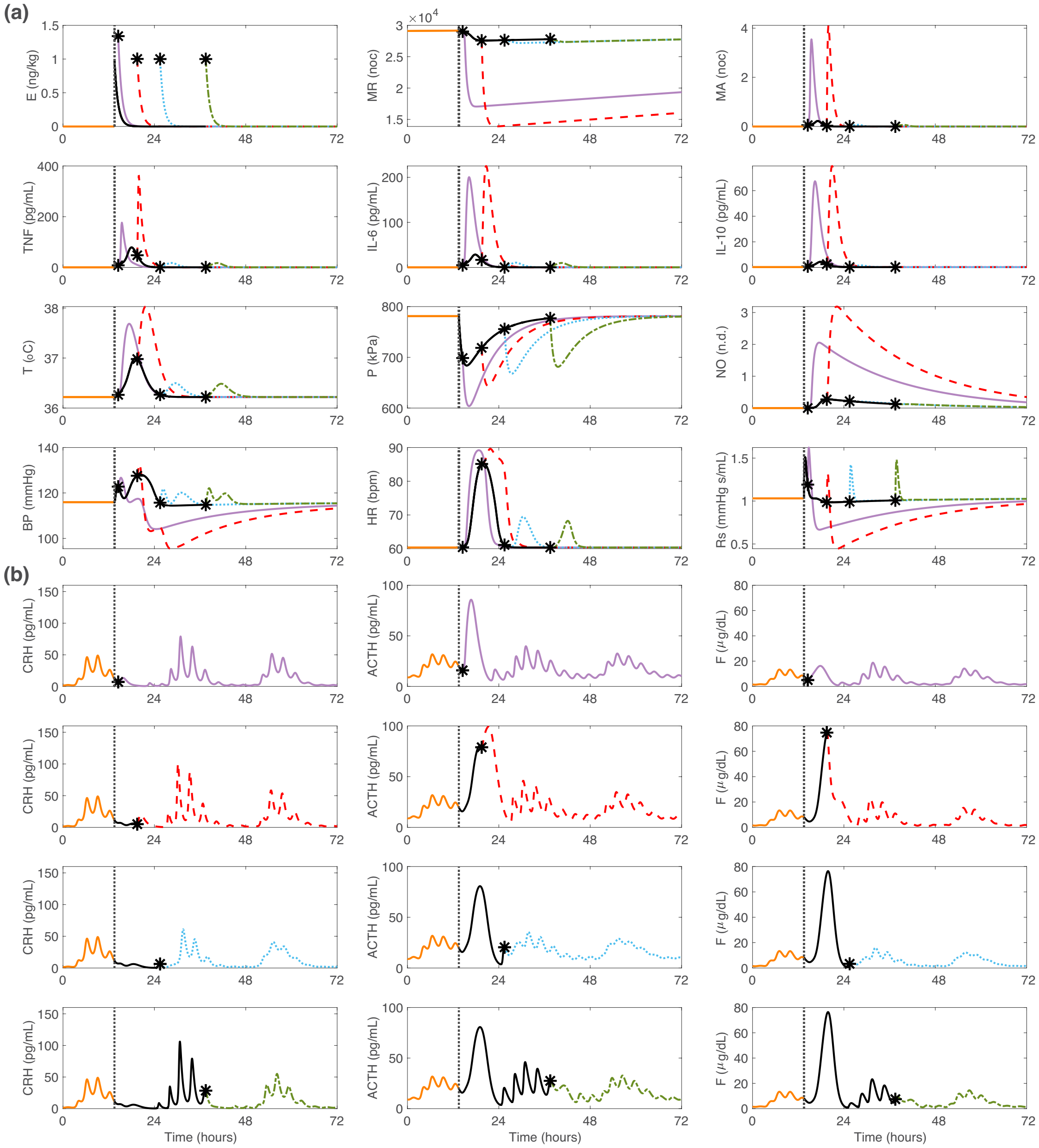

**Fig. 9** Repeated LPS dose: Model simulations when 1 ng/kg is administered at $t_1 = 13.5$ hours and 1 ng/kg is administered at $t_2$. The solid orange line denotes the model simulation before the first LPS dose. The vertical dotted line denotes when the first LPS dose is given ($t_1 = 13.5$). The solid black line denotes the model dynamics between the first and second LPS doses. Black asterisks denote when the second LPS dose is given. The second LPS dose is administered at times $t_2 = 14.5, 19.5, 25.5,$ and $37.5$ hours. Model dynamics after the second dose of LPS is administered are given by purple solid lines ($t_2 = 14.5$), red dashed lines ($t_2 = 19.5$), blue dotted lines ($t_2 = 25.5$), and green dashed-dotted lines ($t_2 = 37.5$). (**a**) Inflammatory and cardiovascular simulations. (**b**) HPA axis simulations.

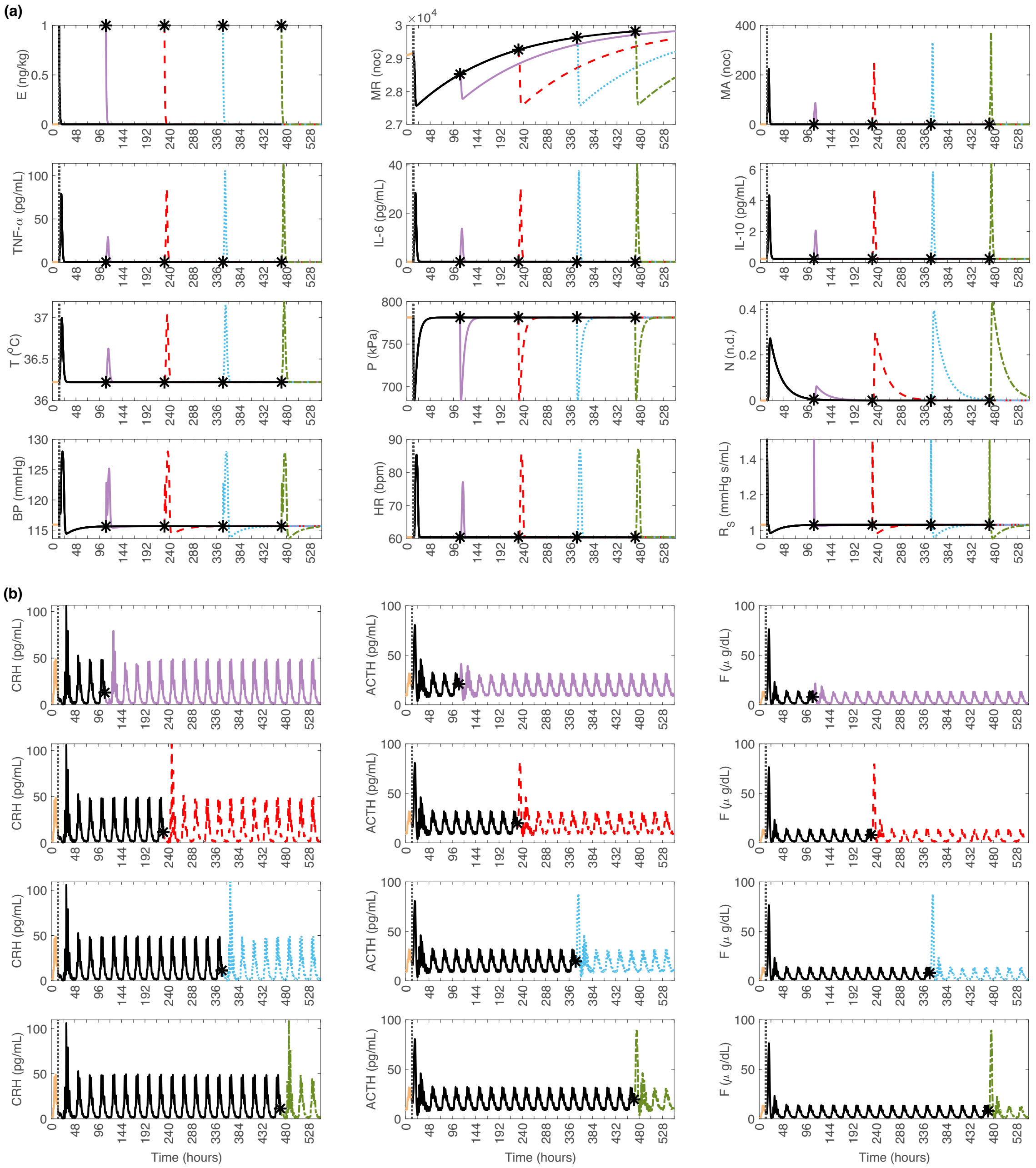

**Fig. 10** Repeated LPS dose further investigation: Model simulations when 1 ng/kg is administered at $t_1 = 13.5$ hours and 1 ng/kg is administered at $t_2$. The solid orange line denotes the model simulation before the first LPS dose. The vertical dotted line denotes when the first LPS dose is given ($t_1 = 13.5$). The solid black line denotes the model dynamics between the first and second LPS doses. Black asterisks denote when the second LPS dose is given. The second LPS dose is administered 4, 9, 14, and 19 24-hour cycles following the first LPS dose. Model dynamics after the second dose of LPS is administered are given by purple solid lines ($t_2 = 120.5$), red dashed lines ($t_2 = 240.5$), blue dotted lines ($t_2 = 360.5$), and green dashed-dotted lines ($t_2 = 480.5$). (**a**) Model simulations for non-HPA axis states. (**b**) Model simulations for HPA axis states.

These results demonstrate that effects persist even if the second dose is administered 24 hours after the first dose. Therefore we added additional simulations, administering the second dose 4, 9, 14, and 19 24-hour cycles following the first LPS dose (shown in Figure 10). These results allow us to investigate how long it will take before the effect of the first dose has worn off and when the impact of the second dose is similar to that of the first dose. The simulation results shown in Figure 10 demonstrate that the system takes approximately nine 24-hour cycles to repeat the dynamics from the first LPS dose. Cytokine levels, body temperature, pain perception threshold, nitric oxide, and cardiovascular markers in Figure 10**(a)** replicate the system's initial response to 1 ng/kg of LPS after nine 24-hour cycles. For the HPA states, we see approximately the same response as with the initial 1 ng/kg dose when the second dose is administered about nine 24-hour cycles later. We also see a more pronounced response in the system after nine 24-hour cycles because the resting monocyte population is being regenerated to a larger monocyte pool than when the first dose was administered (populating up to its carrying capacity $M_\infty$). Therefore, more activated monocytes enter the system and produce an increased inflammatory response which cascades to other model states.

## 3.4 Effect of dose in single LPS injection

The model described in this paper is calibrated to data from studies administering a 2 ng/kg bolus dose of LPS. However, the stimulation of the system is much stronger when diseases such as sepsis elicit an immune response. Experimentally, substantially higher doses of LPS are not safe, but any dose can be administered computationally. Therefore, we investigate the system dynamics for higher LPS doses. To ensure that we stay within the region for which the model is developed, we study the effect of 2, 4, 8, and 16 ng/kg bolus injections. Figure 11 shows the simulation results.

More monocytes are activated as the dose increases. The larger LPS doses cause a slight increase in TNF-$\alpha$ levels and a significant increase in IL-6 and IL-10 levels. Increases occur faster with the larger doses, and TNF-$\alpha$ returns to baseline faster. For the HPA states, the short-term CRH response is suppressed for larger LPS doses. Moreover, the next cycle is slightly affected. We also note that the immediate ACTH peak is not affected, but the next cycle is. Cortisol has a short-term peak response for low doses but is suppressed for larger LPS doses. However, the next cortisol cycle is markedly affected only for low doses.

The cardiovascular state blood pressure and vascular resistance show more prominent initial peaks and then plunges below 100 mmHg for the higher doses, causing hypotension. The recovery time is not affected by the dose. We also note that the resistance falls slightly more with the higher doses, while the peak heart rate response decreases and returns to baseline faster. Finally, the nitric oxide level increases with approximately the same recovery time. For the remaining states, peak temperature slightly increases with the larger doses, and the pain perception threshold dramatically decreases, but with the same recovery time.

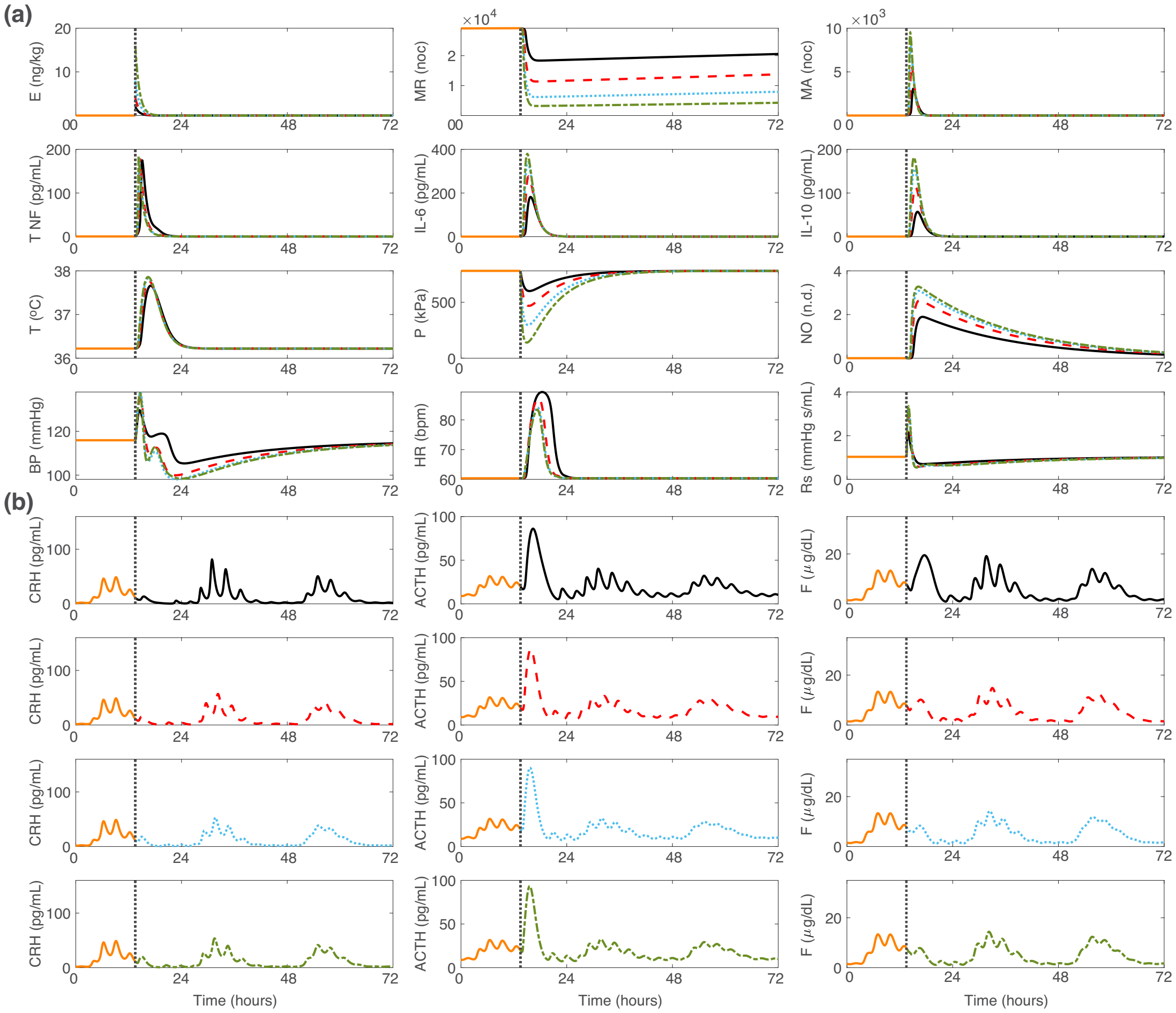

**Fig. 11** Model simulations when the total amount of LPS administered as a bolus dose varies. The solid orange line denotes the model simulation before LPS administration. The vertical dotted line indicates when the LPS was given ($t$ = 13.5 hours). Post-LPS model dynamics are shown by solid black lines (2 ng/kg), red dashed lines (4 ng/kg), blue dotted lines (8 ng/kg), and green dashed-dotted lines (16 ng/kg). (**a**) Inflammatory and cardiovascular simulations. (**b**) HPA axis simulations.

### 3.5 Continuous LPS administration

When endotoxin (LPS), an immuno-stimulant, is injected into the body, it elicits an immune response, eliminating the endotoxin from the system. Studies have suggested that a single bolus dose is insufficient for mimicking realistic inflammatory reactions seen during systemic inflammation as it does not consider that immuno-stimulants, whether from an infection or an injury, aggravate the system for an extended period [49, 58, 77, 98]. Therefore, we examine the system's response to a continuous LPS infusion using the same total dosage (2 ng/kg). In our view, this is a better representation of the inflammatory response in diseases such as sepsis.

The continuous infusion is simulated using the second branch of the endotoxin equation (1). The initial dose, $E(0)$, is set to zero and we include a constant production over a set time interval $t_{cd}$. This is combined with exponential decay, which eliminates the endotoxin from the system. We apply a continuous dose over four hours, making the administered concentration 0.5 ng/kg per hour in the body. The model response is shown in Figure 12.

As expected, the endotoxin concentration changes from a decreasing exponential curve rapidly approaching zero to a distributed reaction to the LPS infusion approaching 0.5 ng/kg. For the inflammation states, the response to LPS is delayed. Moreover, peak values are higher than those observed for the single dose, particularly the late pro- and anti-inflammatory states. After end-infusion, the cytokine states return to their baseline after about 10 hours. The monocyte response is delayed, and the drop in resting monocytes and the corresponding peak in activated monocytes is pronounced after end-infusion. The monocytes approach the baseline level with the same speed as after the single bolus simulation.

For the HPA states, the ultradian CRH peaks reach lower values, and the frequency is slightly increased compared to a single bolus administration. The ACTH response is similar to that of a single bolus, but it is delayed with a change in ultradian frequency while the peak value remains unchanged. The initial cortisol reaction is more suppressed and delayed than for the single bolus simulation, while the ultradian frequency and amplitudes are like those for CRH.

The cardiovascular states behave similarly to the single dose, except the peak value is delayed for all states. Blood pressure changes more than the other states. Its peak is lower, and it drops significantly below 100 mmHg. This trend is repeated for resistance, increasing less and dropping more than for a single dose. NO production is increased, but it returns to baseline at the same time as the bolus dose. The pain perception threshold also returns to baseline at the same time as for a single dose.

## 4 Discussion

We have developed a unified model integrating submodels for the immune system, the endocrine HPA axis, the cardiovascular system, temperature, and pain perception. The unified model integrates complex dynamic features acting on multiple timescales. The novelty involves the development of a platform for studying how coupled submodels impact dynamics over 24 hours. Previous studies have examined how stress impacts inflammation [6], and how inflammation impacts cardiovascular dynamics, temperature, and pain [29, 32, 90, 64, 109]. Still, to our knowledge, this is the first study coupling all five systems to examine the response to endotoxin.

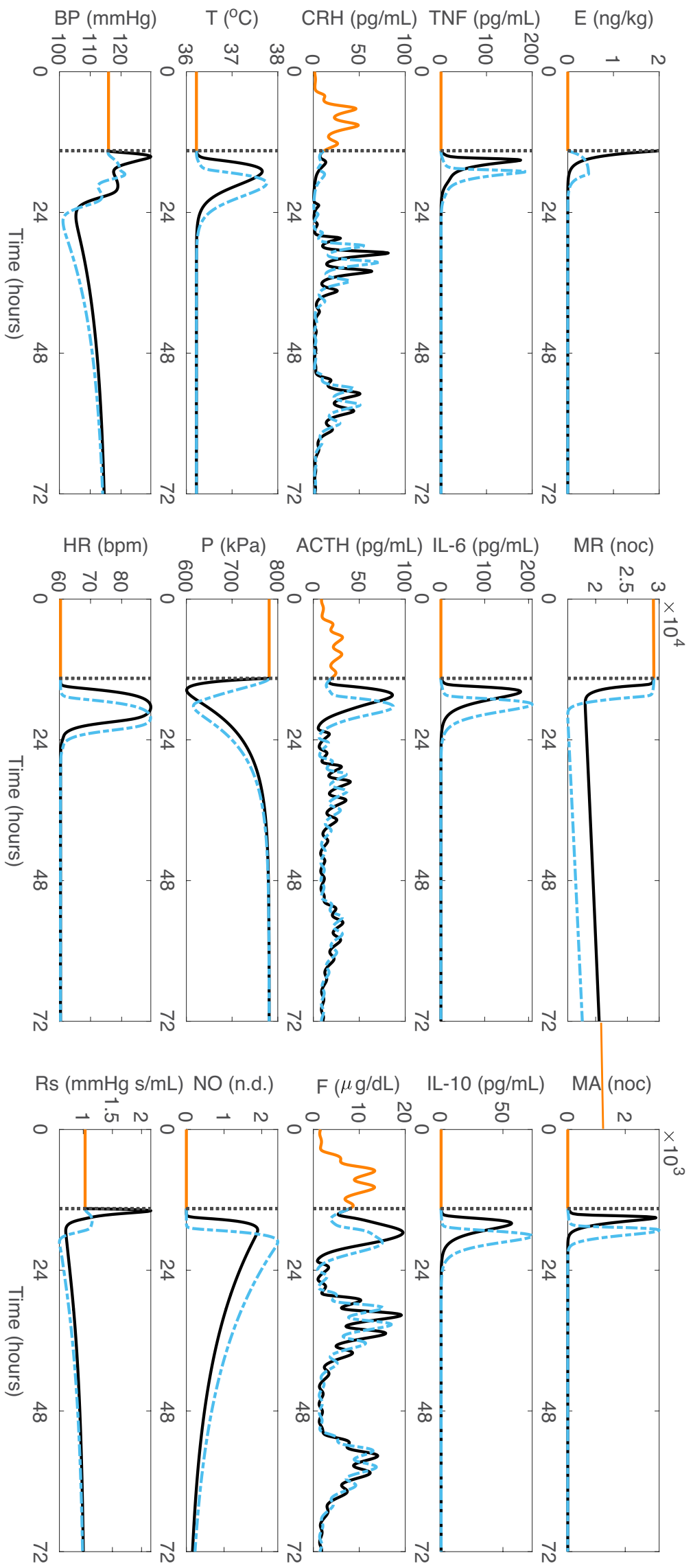

**Fig. 12** Model simulations for continuous LPS administration. We simulate a 2 ng/kg infusion over 4 hours. The solid orange lines denote results before LPS administration. The vertical dotted lines mark when the LPS is given (at $t = 13.5$ hours). The solid black lines show results for a 2 ng/kg bolus injection, and the dotted blue lines show results with continuous infusion.

Mathematical models analyzing physiological systems are typically studied in isolation as it is challenging to parameterize and calibrate an integrated model [46]. The same applies to many *in vitro* experimental studies [95]. Since the ultimate goal is to understand *in vivo* dynamics, more work is needed to develop coupled models.

This study examines the human responses to an endotoxin challenge *in silico*. Specifically, we look at the effect of injected or infused LPS doses, which is also a typical experimental test setup [4, 11]. The advantage is that the LPS effects are relatively short-lived [33] and tolerated by humans and animals [91], making it an ideal controlled environment to learn about system dynamics. Moreover, numerous experimental results provide data we can use to calibrate the proposed model [20, 42].

We focus on examining dynamics up to 48 hours past LPS administration, varying the dose, timing, and administration method. A single bolus dose results in a relatively fast cardiovascular and inflammatory response independent of the administration time, while the HPA-axis hormones are sensitive to administration time. ACTH and cortisol rapidly increase after dispensing LPS, and the next circadian cycle is perturbed. CRH is also susceptible to the administration time; the effect is most pronounced when the administration coincides with the lowest cortisol levels. These results reveal that it takes more than 24 hours before the impact of LPS is cleared from the system. This is important as most *in vivo* LPS experiments only examine the response over 6-8 hours [20, 23, 33, 42]. However, a few recent studies have examined feedback over longer timescales that display results up to 72 hours, but without accounting for circadian or ultradian rhythms [11, 27]. A few recent studies have examined the effects of the circadian clock [48, 81], but these do not address how inflammation interacts with the other subsystems studied here. These results demonstrate that to compare results among subjects, it is essential to conduct experiments at the same time during the day [48, 81].

If the first bolus injection is followed by a second (repeated) injection, our simulations show that if the repeated dose is given within 6 hours, the effect is amplified; but if it is given later, the response is suppressed. Our additional investigation where the second dose was given several 24-hour cycles following the first LPS dose shows that it takes the system approximately ten 24-hour cycles after the initial dose to act as a new independent dose. This emphasizes our earlier point that the initial effects of an LPS dose are present in the body longer than 24 hours. This finding agrees with results reported by Patel et al. [78] examining monocyte regeneration in a study administering 2 ng/kg LPS. Their results show that by day 7, monocyte numbers had returned to steady-state values.

The effect of the bolus injection is mainly dependent on the size of the dose. The higher the dose, the higher the response, except for heart rate, CRH, and cortisol, which all show an opposite dependence. As discussed in our previous studies [10, 29], this model does not include a tissue damage component. Therefore, we do not observe a phase transition as reported in several previous modeling studies [18, 96] or the recent review by Reynolds et al. [84]. While tissue damage is essential for simulating an infection caused by a wound or surgery, it is not relevant for a controlled LPS study. Typical doses given to humans are between 2-4 ng/kg [11], and it has been reported that doses of 1-2 $\mu$g [28] are lethal. Therefore, we limited the dose to 16 ng/kg, significantly above the limit administered in controlled trials but well below the lethal limit. Another consideration was administering a dose for which the physiological pathways included in the model still are valid. Our results show that for 16 ng/kg, blood pressure dropped to ~100 mmHg, which is approaching clinically-defined hypotension. More work is needed to test if a higher dose puts the response below this limit.

A continuous infusion over four hours with the same total LPS dose as in the 2 ng/kg single bolus dose shows distributed and amplified effects compared to a single bolus. Exceptions are pain perception and cortisol, where the outcome is less than for a single dose, as well as heart rate and ACTH, where the effect is comparable. For CRH, ACTH, and cortisol, the ultradian frequency increases for the continuous infusion compared to the single bolus injection. Again, the longer infusion time causes a significant drop in blood pressure. Still, more work is needed to examine if the same dose administered over a longer time would cause a higher blood pressure drop or if it is necessary also to increase the total dose.

Results reported here bring new insight into how the systems couple. The model proposed here includes primary coupling between five subsystems. While numerous studies have examined parts of these [6, 10, 29, 32, 90, 64, 109], to our knowledge, this pilot study devising a unified model is the first to explore connections between these subsystems.

**Limitations.** This study devised a unified model calibrated to data from low dose LPS injections [20, 42]. Results were obtained by combining submodels from our previous studies [6, 29]. This approach has two limitations (1) we do not have a single data set measuring all states, and (2) coupling between the two models can be challenging to validate. Since we did not have a single set of data, we calibrated our model to the cytokine, hormonal, and temporal data from [20] and the cardiovascular and pain data from [42]. However, it should be noted that cytokine data could be taken from either study since it was measured in both studies. Therefore, more work is needed to set up experimental studies measuring all required quantities in response to a single LPS dose and examine how the coupling strength between each submodel impacts dynamics.

Moreover, even though each submodel is "simple," (we only include two pro- and one anti-inflammatory cytokine, three HPA-axis hormones, and four compartments in the systemic circulation), the combined unified model is highly complex. It has 18 state variables and more than 100 parameters. Therefore, in addition to model calibration against data from a single experiment, more work is needed to study the sensitivity and uncertainty of model predictions.

Sensitivity analysis can reveal what parameters are the most influential, and if combined with subset selection, it is possible to determine what parameters can be identified uniquely given available data. The latter is advantageous for improving model validation. This analysis should focus on detailed scrutiny of parameters informing coupling between each submodel, mainly since each submodel has been analyzed in detail in earlier studies [6, 10, 29]. Finally, adding uncertainty quantification could benefit predictions to relate parameters to significant variation often observed in experimental measurements of dynamic quantities studied here.

However, more work is needed to examine the response to other types of infections, mainly since LPS mimics a bacterial infection that does not replicate in the body. Therefore, results reported here noting that it takes more than 48 hours before the LPS has cleared should be scrutinized if the model is generalized to investigate the effects of fungal or viral infections. The same applies to other types of infection, including inflammation associated with surgery, systemic infection like sepsis, chronic inflammation from autoimmune diseases, the mutation in cells giving rise to a pathogenic reaction, smoking, aging, or other immune modulators.

## 5 Conclusion

This study develops a unified model coupling inflammatory, HPA axis, and cardiovascular dynamics. Results show that infection generated by administration of low dose endotoxin (LPS) takes at least 48 hours to clear and that dosing type, amount, and timing affect dynamics. In particular, the repeated dose has a significant effect demonstrating that it takes about 10 days before the second dose is not influenced by the first. Another important finding is that several manipulations cause a substantial drop in blood pressure towards hypotension, a significant risk for patients. This effect is seen if the dose is administered over a longer time via a repeated injection or a continuous infusion, or if an increase in endotoxin dose is given.

## Appendix

Table 1: Model parameters and initial conditions. Parameters with reference ~ were scaled from their values reported in [6] and [29] to match the appropriate variable concentration in the model, parameters with a * indicates that the parameter was manually adjusted , and parameters with ~* were both scaled and manually adjusted.

| Parameter | Meaning | Value | Unit | Reference |
|---|---|---|---|---|
| | Inflammation model | | | |
| $k_E$ | Endotoxin decay rate | 1.08 | $hr^{-1}$ | [29] |
| $k_M$ | $E$ activation rate of monocytes | 0.0414 | $hr^{-1}$ | [29] |
| $k_{MTNF}$ | TNF-$\alpha$ activation rate of monocytes | 8.65 | $hr^{-1}$ | [29] |
| $k_{MR}$ | $M_R$ regeneration rate | $6\times10^{-3}$ | $hr^{-1}$ | [29] |
| $M_\infty$ | Monocyte carrying capacity | $3\times10^4$ | noc | [29] |
| $k_{MA}$ | $M_A$ decay rate | 2.51 | $hr^{-1}$ | [29] |
| $\eta_{ME}$ | Half-max of $E$ upreg $M_A$ | 3.3 | ng/kg | [29] |
| $\eta_{M10}$ | Half-max of IL-10 downreg $M_A$ | 4.35 | pg/mL | [29] |
| $\eta_{MTNF}$ | Half-max of TNF-$\alpha$ upreg $M_A$ | 41.7 | pg/mL | ~ |
| $h_{ME}$ | Exp. of $E$ upregulation of $M_A$ | 1 | n.d. | [29] |
| $h_{M10}$ | Exp. of IL-10 downregulation of $M_A$ | 0.3 | n.d. | [29] |
| $h_{MTNF}$ | Exp. of TNF-$\alpha$ upreg $M_A$ | 3.16 | n.d. | [29] |
| $k_{TNFM}$ | Rate of $M_A$ production of TNF-$\alpha$ | 0.290 | pg/(mL hr noc) | ~ |
| $k_{TNFHR}$ | HR effect on TNF-$\alpha$ production | 0.05 | $(\text{bpm})^{-1}$ | * |
| $k_{TNF}$ | TNF-$\alpha$ decay rate | 1 | $\text{hr}^{-1}$ | * |
| $w_{TNF}$ | TNF-$\alpha$ baseline amount | 0.466 | pg/mL | ~ * |
| $\eta_{TNF10}$ | Half-max of IL-10 downreg TNF-$\alpha$ | 17.4 | pg/mL | [29] |
| $\eta_{TNF6}$ | Half-max of IL-6 downreg TNF-$\alpha$ | 140 | pg/mL | ~ |
| $h_{TNF10}$ | Exp. of IL-10 downreg TNF-$\alpha$ | 3 | n.d. | [29] |
| $h_{TNF6}$ | Exp. of IL-6 downreg TNF-$\alpha$ | 2 | n.d | [29] |
| $k_{6M}$ | Rate of IL-6 production due to $M_A$ | 0.226 | pg/(mL hr noc) | ~ * |
| $k_{6TNF}$ | Rate of TNF-$\alpha$ production of IL-6 | 0.253 | pg/(mL hr noc) | ~ * |
| $k_6$ | IL-6 decay rate | 0.797 | $\text{hr}^{-1}$ | [29] |
| $w_6$ | IL-6 baseline amount | 0.262 | pg/mL | ~ * |
| $\eta_{610}$ | Half-max of IL-10 downreg IL-6 | 34.8 | pg/mL | [29] |
| $\eta_{66}$ | Half-max of IL-6 downreg IL-6 | 140 | pg/mL | ~ |
| $\eta_{6TNF}$ | Half-max of TNF-$\alpha$ upreg IL-6 | 77.1 | pg/mL | ~ |
| $h_{610}$ | Exp. of IL-10 downreg | 0.25 | n.d. | * |
| $h_{66}$ | Exp. of IL-6 downreg IL-6 | 0.25 | n.d. | * |
| $h_{6TNF}$ | Exp. of TNF-$\alpha$ upreg IL-6 | 0.25 | n.d. | * |
| $k_{10M}$ | Rate of IL-10 production due to $M_A$ | 0.0105 | pg/(mL hr noc) | * |
| $k_{106}$ | Rate of IL-10 production due to IL-6 | 0.0191 | pg/(ml hr noc) | [29] |
| $k_{10F}$ | Rate of IL-10 production due to $F$ | 0.01 | pg/(mL noc hr) | * |
| $k_{10}$ | IL-10 decay rate | 0.834 | $\text{hr}^{-1}$ | * |
| $w_{10}$ | IL-10 baseline amount | 0.235 | pg/mL | * |
| $\eta_{106}$ | Half-max of IL-6 upreg IL-10 | 140 | pg/mL | ~ |
| $\eta_{10F}$ | Half-max of $F$ upreg IL-10 | 0.8 | pg/mL | * |
| $h_{106}$ | Exp. of IL-6 upreg IL-10 | 3.68 | n.d. | [29] |
| $h_{10F}$ | Exp. of $F$ upreg IL-10 | 10 | n.d. | * |
| | | HPA axis model | | |
| $k_{CR}$ | Cir. rhy. rate of $C$ production | $4.10\times10^{11}$ | $\text{hr}^{-1}$ | [6] |
| $\eta_{CF}$ | Half-max of $F$ downreg $C$ | $2.39\times10^{-5}$ | $\mu$g/dL | [6] |
| $h_{CF}$ | Exp. of $F$ downreg $C$ | 2 | n.d. | [6] |
| $k_{CTNF}$ | Rate of $C$ production by TNF-$\alpha$ | 0.160 | $\text{hr}^{-1}$ | ~ |
| $k_C$ | $C$ decay rate | 1.92 | $\text{hr}^{-1}$ | [6] |
| $C_b$ | Baseline $C$ level | 0.06 | pg/mL | [6] |
| $k_{AC}$ | Rate of $A$ production by $C$ | $1.42\times10^6$ | $\text{hr}^{-1}$ | [6] |
| $\eta_{FA}$ | Half-max of $F$ downreg $A$ | $5.62\times10^{-6}$ | pg/mL | [6] |
| $h_{FA}$ | Exp. of $F$ downreg $A$ | 1 | n.d. | * |
| $k_{ATNF}$ | Rate of $A$ production by TNF-$\alpha$ | 100 | pg/(mL hr) | * |
| $\eta_{ATNF}$ | Half-max of TNF-$\alpha$ upreg $A$ | 33.33 | pg/mL | ~ |
| $h_{ATNF}$ | Exp. of TNF-$\alpha$ upreg $A$ | 2 | n.d. | [6] |
| $k_A$ | $A$ decay rate | 0.96 | $\text{hr}^{-1}$ | [6] |
| $k_{FA}$ | Rate of $F$ production by $A$ | 0.0255 | $\mu$g mL/(pg dL hr) | * |
| $\eta_{F10}$ | Half-max of IL-10 downreg $F$ | 10 | pg/mL | * |
| $h_{F10}$ | Exp. of IL-10 downreg $F$ | 1 | n.d. | * |
| $k_F$ | $F$ decay rate | 1.56 | $\text{hr}^{-1}$ | * |
| | Circadian Rhythm | | | |
| $\alpha$ | Upregulation half-max | 300 | min | [6] |
| $k$ | Upregulation exp. | 5 | n.d. | [6] |
| $\beta$ | Downregulation half-max | 950 | min | [6] |
| $\ell$ | Downregulation exp. at $t = \beta$ | 6 | n.d. | [6] |
| $\varepsilon$ | Baseline circadian rhythm level | 0.01 | n.d. | [6] |
| $N_c$ | Scaling factor | 1.92 | n.d. | * |
| $\delta$ | Circadian clock time shift | 70 | min | * |
| | Cardiovascular circulation model | | | |
| $R_a$ | Arterial resistance | 686 | mmHg hr/mL | * |
| $R_{sb}$ | Baseline peripheral resistance | 3713 | mmHg hr/mL | * |
| $R_v$ | Venous resistance | 9.72 | mmHg hr/mL | * |
| $E_{la}$ | Large artery elastance | 0.791 | mmHg/ mL | * |
| $E_{sa}$ | Small artery elastance | 3.92 | mmHg/ mL | * |
| $E_{sv}$ | Small vein elastance | 0.132 | mmHg/mL | * |
| $E_{lv}$ | Large vein elastance | 0.0217 | mmHg/mL | * |
| $E_{Max}$ | Maximum elastance | 3.20 | mmHg/mL | * |
| $E_{Min}$ | Minimum elastance | 0.0265 | mmHg/mL | * |
| | Cardiovascular control model | | | |
| $k_H$ | Rate of change of HR | 0.25 | n.d. | * |
| $\tau_H$ | HR time constant | 0.791 | hr | * |
| $H_M$ | Maximum HR | 190 | bpm | [35] [42] |
| $H_b$ | Baseline HR | 60.4 | bpm | [42] |
| $p_{lab}$ | Baseline BP | 118 | mmHg | [42] |
| $\eta_{HT}$ | Half-max of $T$ upreg HR | 0.354 | °C | * |
| $h_{HT}$ | Exp. of $T$ upreg HR | 2 | n.d. | [29] |
| $\eta_{HF}$ | Half-max of $F$ upreg HR | 5 | $\mu$g/dL | * |
| $h_{HF}$ | Exp. of $F$ upreg HR | 4 | n.d. | * |
| $\eta_{Hp}$ | Half-max of BP regulating HR | 24.6 | mmHg | [29] |
| $h_{Hp}$ | Exp. of BP regulating HR | 4 | n.d. | [29] |
| $k_{NM}$ | $M_A$ production rate of NO | 0.002 | $(\text{hr noc})^{-1}$ | [29] |
| $k_N$ | NO decay rate | 0.045 | $\text{hr}^{-1}$ | [29] |
| $\eta_{NTNF}$ | Half-max of TNF-$\alpha$ upreg NO | 39.6 | pg/mL | ~ |
| $\eta_{N10}$ | Half-max of IL-10 downreg NO | 4 | pg/mL | [29] |
| $h_{NTNF}$ | Exp. of TNF-$\alpha$ upreg NO | 2 | n.d. | [29] |
| $h_{N10}$ | Exp. of IL-10 downreg NO | 0.4 | n.d. | [29] |
| $k_{RP}$ | Rate of $R_S$ stimulation by $P$ | 13.0 | mmHg/mL | * |
| $k_{RN}$ | Rate of $R_S$ inhibition by NO | 0.8 | mmHg/mL | * |
| $k_R$ | $R_S$ recovery rate | 4.28 | $hr^{-1}$ | * |
| $\eta_{RP}$ | Half-max of $P$ upreg $R_S$ | 230 | kPa/hr | [29] |
| $h_{RP}$ | Exp. of $P$ upreg $R_S$ | 2 | n.d. | [29] |
| | **Pain model** | | | |
| $k_{PE}$ | Rate of change of $P$ due to $E$ | 0.2 | kg/(ng hr) | [29] |
| $k_P$ | Rate of $P$ recovery | 0.15 | $hr^{-1}$ | [29] |
| $P_b$ | Baseline $P$ level | 781 | kPa | [42] |
| | **Temperature model** | | | |
| $T_b$ | Baseline temperature | 36.2 | °C | [20] |
| $k_T$ | $T$ recovery rate | 0.7 | $hr^{-1}$ | * |
| $k_{TTNF}$ | Rate of $T$ upregulation by TNF-$\alpha$ | 1 | °C/hr | * |
| $k_{T6}$ | Rate of $T$ upregulation by IL-6 | 1.9 | °C/hr | * |
| $k_{T10}$ | Rate of $T$ downregulation by IL-10 | 0.2 | °C/hr | * |
| $\eta_{TTNF}$ | Half-max of TNF-$\alpha$ upreg $T$ | 130 | pg/mL | ~ * |
| $\eta_{T6}$ | Half-max of IL-6 upreg $T$ | 140 | pg/mL | ~ |
| $\eta_{T10}$ | Half-max of IL-10 downreg $T$ | 40 | pg/mL | * |
| $h_{TTNF}$ | Exp. of TNF-$\alpha$ upreg $T$ | 1 | n.d. | * |
| $h_{T6}$ | Exp. of IL-6 upreg $T$ | 1 | n.d. | * |
| $h_{T10}$ | Exp. of IL-10 downreg $T$ | 1 | n.d. | [29] |

**Initial Conditions**

| Variable | Model State | Value | Unit |
|---|---|---|---|
| $E$ | Endotoxin | 0 | ng/kg |
| $M_R$ | Resting monocytes | 28200 | noc |
| $M_A$ | Activated monocytes | 0 | noc |
| $TNF$ | TNF-$\alpha$ | 0.466 | pg/mL |
| $IL6$ | IL-6 | 0.262 | pg/mL |
| $IL10$ | IL-10 | 0.234 | pg/mL |
| $C$ | CRH | 1.368 | pg/mL |
| $A$ | ACTH | 8.872 | pg/mL |
| $F$ | Cortisol | 1.590 | $\mu$g/dL |
| $V_{la}$ | Large artery volume | 146.647 | mL |
| $V_{sa}$ | Small artery volume | 25.876 | mL |
| $V_{sv}$ | Small vein volume | 28.162 | mL |
| $V_{lv}$ | Large vein volume | 159.943 | mL |
| $HR$ | Heart rate | 60.335 | bpm |
| $R_S$ | Vascular resistance | 1.040 | mmHg hr/mL |
| $N$ | Nitric oxide | 0 | n.d. |
| $P$ | Pain perception threshold | 781.5 | kPa |
| $T$ | Temperature | 36.22 | °C |

upreg = upregulating, downreg = downregulating, exp = exponent